\documentclass[reprint,aps,pre,floatfix,onecolumn,fleqn,longbibliography,superscriptaddress]{revtex4-2}
\usepackage{amsmath,amssymb}
\usepackage{graphicx}
\usepackage{natbib}

\begin{document}
\preprint{APS/123-QED}

%\title{Erosion and recovery of synchronization among noisy oscillators with simplicial interactions}
%\title{Erosion of synchronization and its prevention among noisy oscillators with simplicial interactions}
% \title{Synchronization and desynchronization in noisy oscillators with simplicial interactions}
\title{Synchronization and its slow decay in noisy oscillators with simplicial interactions}
\author{Yuichiro Marui}
\email[]{yuichiromarui@gmail.com}
\affiliation{Department of Mathematical Informatics, Graduate School of Information Science and Technology, The University of Tokyo, 113-8656, Japan}
\author{Hiroshi Kori}
\email[]{kori@k.u-tokyo.ac.jp}
% \affiliation{Department of Mathematical Informatics, Graduate School of Information Science and Technology, The University of Tokyo, 113-8565}
\affiliation{Department of Complexity Science and Engineering, The University of Tokyo, 277-8561, Japan}
\date{\today}

\begin{abstract}
Previous studies on oscillator populations with two-simplex interaction have reported novel phenomena such as discontinuous desynchronization transitions and multistability of synchronized states. However, the noise effect is not well understood.
Here, we study a higher-order network of noisy oscillators with generic interactions consisting of one-simplex and two types of two-simplex interactions. We observe that 
%Here, we find that when oscillators with two-simplex interaction
%alone
when a type of two-simplex interaction is dominant, synchrony is eroded and eventually disappears even for infinitesimally weak noise.
%when the noise is infinitesimally weak.
%and weak or absent one-simplex interaction are subjected to external noise, 
Nevertheless, synchronized states may persist for extended periods, with the lifetime increasing approximately exponentially with the strength of the two-simplex interaction.
When one-simplex or another type of two-simplex interaction is sufficiently strong, 
%Further, when sufficiently strong additionally introduced,
noise erosion is prevented, and synchronized states become persistent.
A weakly nonlinear analysis reveals that as one-simplex coupling increases, the synchronized state appears supercritically or subscritically, depending on the interaction strength.
Furthermore, assuming weak noise and using Kramers' rate theory, we derive a closed dynamical equation for the Kuramoto order parameter, from which the time scale of the erosion process is derived.
%The bifurcation analysis of the desynchronized state
 Our study elucidates the synchronization and desynchronization of oscillator assemblies in higher-order networks and is expected to provide insights into such systems' design and control principles.
\end{abstract}

\maketitle

\section{Introduction}
% {\em Introduction.}
Synchronization is a major field of study in not only physics but also chemistry, biology, engineering, and sociology \cite{winfree01,kuramoto84,Pikovsky2001-hv,glass01,Eilam2019-eh}.
Examples include
pacemaker cells in the heart \cite{Glass2020-qs}, laser arrays \cite{Simonet1994-rq}, applauding audiences \cite{Neda2000-yv, Neda2000-io}, power grids consisting of alternating current (AC) generators \cite{Rohden2012-gf}, and Josephson junctions
\cite{Wiesenfeld1996-kl, Wiesenfeld1998-qi}.
In addition to synchronization, understanding desynchronization in oscillator assemblies is crucial. For example, 
while synchronization of circadian pacemaker cells in the brain is essential for mammals to maintain 24-hour activity rhythm, their transient desynchronization, triggered by a phase shift of light-dark cycles, is a putative cause of jet lag symptoms \cite{yamaguchi13, kori17}.
Desynchronization is also significant in neurological disorders such as Parkinson's disease \cite{popovych2018multisite} and epilepsy \cite{jiruska2013synchronization}. Methods for promoting desynchronization in this context have been actively studied both theoretically and experimentally \cite{pikovsky2015dynamics}.

Phase oscillator models, including Kuramoto's model \cite{Kuramoto1975-qj}, are widely recognized for their utility, not only in understanding synchronization \cite{kuramoto84,pikovsky01} but also in controlling real-world systems \cite{kiss07,kori08}.
While the classical Kuramoto model considers pairwise interactions between oscillators, recent studies have extended the model to allow for non-pairwise interactions \cite{Tanaka2011-zo,Skardal2019-br,Skardal2020-wq,Xu2020-qk}.
Such structures are often called simplexes, where $n$-simplex describes an interaction between $n + 1$ oscillators \cite{Salnikov2018-jg}. 
Research on brain dynamics \cite{Giusti2016-ks,Sizemore2018-jw} or social phenomena \cite{Wang2020-up} has suggested that simplicial structures play an important role in such systems.

In noiseless phase oscillators with two-simplex interactions,
Tanaka and Aoyagi \cite{Tanaka2011-zo} noted that multiple stable synchronized states appear as two clusters with different population ratios.
Moreover, Skardal and Arenas showed that abrupt desynchronization transitions occur as the interaction strength decreases \cite{Skardal2019-br}.
Skardal and Arenas reported that two- and three-simplicial couplings promote abrupt synchronization transitions in the presence of one-simplex interactions \cite{Skardal2020-wq}.
In addition, numerous relevant studies on the synchronization of noiseless oscillators in higher-order networks have been conducted recently
\cite{Millan2020-xf, Chutani2021-ub, Kuehn2021-mt, Rajwani2023-yd, Carletti2023-vz}
In contrast, for noisy phase oscillators, Komarov and Pikovsky \cite{Komarov2015-es}
studied a system with a particular type of two-simplicial coupling alone
and reported that no stable synchronized states exist in the limit of an infinite 
number of oscillators. They mainly focused on the synchronized states that exist only in small populations.
%Desynchronization is expected in a large population, and as mentioned above, understanding its process is essential. 
Desynchronization is expected in a large population, and as mentioned above, clarifying its process is essential. 

In the present study, we consider a large population of noisy phase oscillators with one-simplex and two types of two-simplex couplings.
Although steady synchronized states do not exist when a type of two-simplex interaction, denoted as type-a coupling, is dominant,
%in the system that Komarov and Pikovsky \cite{Komarov2015-es} investigated, 
we demonstrate that the population is transiently synchronized for an extended period and then abruptly desynchronized.
%when two-simplex coupling is sufficiently strong compared to the noise strength. 
Assuming weak noise and exploiting Kramers' rate theory, we derive a closed dynamical equation for the Kuramoto order parameter by which the desynchronization process is reproduced and the exponential dependence of the lifetime of the synchronized states on the type-a coupling strength is derived. 
%We also consider a system in which one- and two-simplex couplings exist and
Further, we show that synchronized states become persistent when one-simplex coupling or another type of two-simplex coupling, denoted as the type-b coupling, is sufficiently strong.
Our bifurcation analysis reveals that the desynchronization-synchronization transition changes from continuous to discontinuous at a critical strength of the type-a coupling.

% % {\em Model and results.}
\section{Model and simulation results}
We consider a system of identical phase oscillators subjected to independent noise and globally coupled with one-simplex (i.e., two-body) and two types of two-simplex (i.e., three-body) interactions, given as
\begin{align}
  \label{eq: model 1}
  \dot{\theta}_m = \omega_m + & \cfrac{K_1}{N} \sum_{j = 1}^N \sin(\theta_j - \theta_m) 
  + \cfrac{1}{N^2} \sum_{j,k = 1}^N \left[K_{2 \mathrm{a}} \sin(\theta_j + \theta_k - 2\theta_m) + K_{2 \mathrm{b}} \sin(2\theta_j - \theta_k - \theta_m)\right]+ \xi_m (t),
\end{align}
where $\omega_m$ and $\theta_m$ are the intrinsic frequency and the phase of the oscillator $m$ ($1 \le m \le N$), respectively. The term $\xi_m(t)$ represents Gaussian white noise with
zero mean, $\delta$-correlated in time and independent for different oscillators. Specifically, 
$\langle \xi_m(t) \rangle = 0,\ \langle \xi_m(t) \xi_n(\tau)\rangle = 2 D \delta_{mn} \delta (t - \tau)$, where $D \ge 0$ is the noise strength.
The remaining terms describe interactions, where $K_1 \ge 0$, $K_{2 \mathrm{a}} \ge 0$, and $K_{2 \mathrm{b}} \ge 0$ are the coupling strengths of one-simplex, type-a two-simplex, and type-b two-simplex interactions, respectively. 
% This model is motivated by the fact that these three types of interactions are generic for oscillators close to the Hopf bifurcation point \cite{ashwin2016hopf}, which
% can be considered a natural extension of the previously proposed models \cite{Komarov2015-es,Skardal2019-br,Skardal2020-wq}. 
This model is motivated by the fact that two types of two-simplicial interaction terms emerge from higher-order phase reductions
\cite{Ashwin2016-xl, Leon2019-qr}, which can be considered a natural extension of the previously proposed models \cite{Komarov2015-es, Skardal2019-br, Skardal2020-wq}.

%Next, we introduce two order parameters defined by
Next, we introduce Kuramoto-Daido order parameters \cite{kuramoto75,daido1996onset}, defined as
\begin{align}
  % Z_l(t) = R_l(t) e^{i\Theta_l(t)} =: \cfrac{1}{N} \sum_{j = 1}^N e^{i l \theta_j(t)}\ \mathrm{for}\ l = 1, 2, 
  Z_l(t) = R_l(t) e^{i\Theta_l(t)} =: \cfrac{1}{N} \sum_{j = 1}^N e^{i l \theta_j(t)},\ \mathrm{for}\ l = \pm 1, \pm 2, \cdots,
\end{align}
where $R_l \in [0, 1]$ and $\Theta_l \in (- \pi, \pi]$ represent the amplitudes and the mean phases, respectively.
%While $Z_1$ is the Kuramoto order parameter, $Z_2$ is the Daido order parameter, which measures synchrony to two clusters separated by phase difference $\pi$. 
Note that $Z_{-l}$ is the complex conjugate of $Z_l$.
We utilize this relationship in Section \ref{subsec: Bifurcation}, where we perform the stability analysis of the incoherent state.
Note that $R_1$ assumes $1$ and $0$ for the in-phase state (i.e., $\theta_j=\theta_0$ for $1 \leq j \leq N$) and the fully desynchronized state (i.e., $\theta_j$ is uniformly distributed within $(- \pi, \pi]$), respectively.
Similarly, $R_2 = 1$ for the two-cluster states given by $\theta_j = \theta_0\ \mathrm{or}\ \theta_0 + \pi$ for $1 \leq j \leq N$ and $R_2 = 0$ for the fully desynchronized state. 
Using $R_l$ and $\Theta_l$, Eq.~\eqref{eq: model 1} may be rewritten as
\begin{align}
  \label{eq: model 1-1}
  \dot{\theta}_m = \omega_m &+ K_1 R_1 \sin(\Theta_1 - \theta_m)
  + K_{2 \mathrm{a}} R_1^2 \sin (2\Theta_1 - 2\theta_m) + K_{2 \mathrm{b}} R_1 R_2 \sin(\Theta_2 - \Theta_1 - \theta_m) + \xi_m (t).
\end{align}
We can observe that
%the three-body interaction term with $K_{2 \mathrm{a}}$ (we refer to this term as the type-a interaction below)
the type-a interaction, given as the third term on the right-hand side, tends to make $\theta_m$ either $\Theta_1$ or $\Theta_1+\pi$. 
%because this force contains the second harmonics. 
%Therefore, one can suspect that the type-a interaction is likely to %promote the formation of two-cluster states.
%This is actually the case for
%$K_1=K_{2 \mathrm{b}}=D=0$ \cite{Skardal2019-br}.
Therefore, one can suspect that the type-a interaction is likely to promote the formation of two-cluster states, which is indeed observed when
$D=0$ \cite{Skardal2019-br}.
Based on this observation, we numerically investigate the dynamics for the initial condition of two-cluster states. 
Specifically, we set $\theta_m(0) = 0$ and $\pi$ for $1 \le m \le \eta N$ and otherwise, respectively, where $\eta \geq \frac{1}{2}$ is the initial population ratio of the two clusters.
Note that $\eta=1$ corresponds to the one-cluster state, i.e., in-phase synchrony.
%We start with

\begin{figure}[htb]
  \begin{tabular}{ll}
    \begin{minipage}[t]{4.5cm}
      \centering
     \includegraphics[height = 4cm]{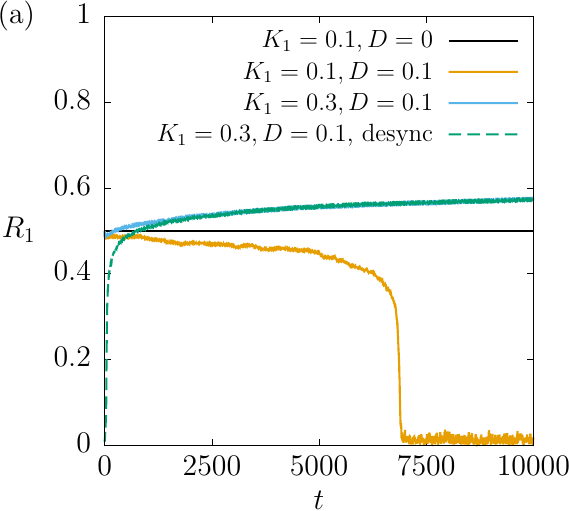}
    \end{minipage} &
    \begin{minipage}[t]{4.5cm}
      \centering
      \includegraphics[height = 4cm]{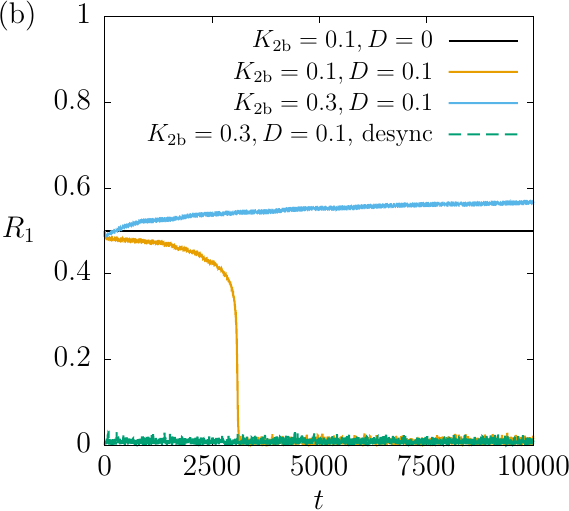}
    \end{minipage}\\
    \begin{minipage}[t]{4.5cm}
      \centering
     \includegraphics[height = 4cm]{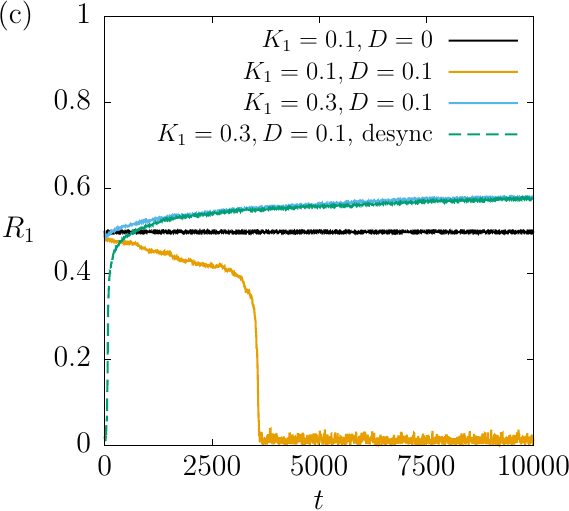}
    \end{minipage} &
    \begin{minipage}[t]{4.5cm}
      \centering
      \includegraphics[height = 4cm]{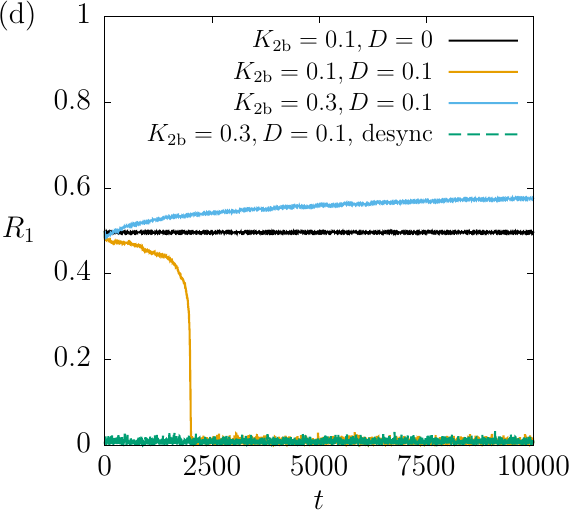}
    \end{minipage}
  \end{tabular}
    \caption{(a, b) Time series of the order parameter $R_1$ for 
 (a) $K_1>0,K_{2 \mathrm{a}}=3.0,K_{2 \mathrm{b}} = 0$, and  (b) $K_1=0, K_{2 \mathrm{a}}=3.0,K_{2 \mathrm{b}} > 0$.
%Type-a coupling strengths are $K_{2 \mathrm{a}} = 3.0$.
We employ the two-cluster state with $\eta=0.75$ for solid curves and the desynchronized state for dashed curves as the initial condition.
(c, d) Time series of the order parameter $R_1$ for (c) $K_1 > 0$, $K_{2 \mathrm{b}} = 0$ and (d) $K_1 = 0$, $K_{2 \mathrm{b}} > 0$
in a population of nonidentical oscillators. 
  Type-a coupling strength is $K_{2 \mathrm{a}} = 3.0$.
  We employ the two-cluster state with $\eta=0.75$ for solid curves and the desynchronized state for dashed curves as the initial condition.
  $N = 10^4$.
} 
    \label{fig: dynamics of R_1}
\end{figure}

\begin{figure}[htb]
  \centering
  \begin{tabular}{ll}
    \begin{minipage}[t]{4.5cm}
      \centering
      \includegraphics[height = 4cm]{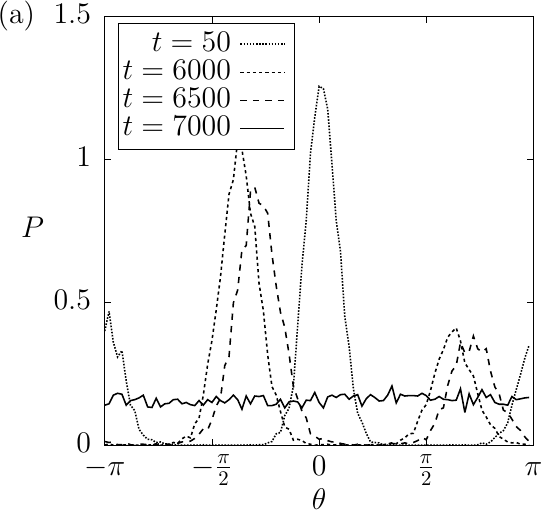}
    \end{minipage}&
    \begin{minipage}[t]{4.5cm}
      \centering
      \includegraphics[height = 4cm]{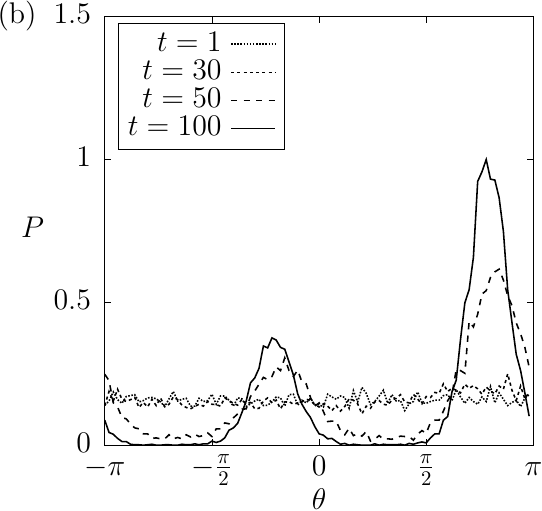}
    \end{minipage}
  \end{tabular}
  \caption{Time courses of the phase distribution $P(\theta,t)$. 
  (a) $K_1=0.1,K_{2 \mathrm{a}}=4.0,K_{2 \mathrm{b}} = 0$
  %$K_1 = 0$, $K_{2 \mathrm{b}} = 0$
  and the initial condition is the two-cluster state with $\eta=0.75$. 
  (b) $K_1=0.3,K_{2 \mathrm{a}}=4.0,K_{2 \mathrm{b}} = 0$
  and the initial condition is the desynchronized state. 
  We obtained the results from simulations of Eq.~\eqref{eq: model 1-1} with $N = 10^4$ oscillators.
}
  \label{fig: distribution}
\end{figure}

We first consider the case of identical oscillators, i.e., $\omega_m = \omega_0$ for $1\leq m \leq N$. Without loss of generality, we set $\omega_0=0$. 
%We further assume $K_{2 \mathrm{b}} = 0$ (i.e. type-b interaction is not present) so as to eliminate the dependence on $R_2$ and $\Theta_2$
%of the governing equation \eqref{eq: model 1-1}. 
%in Eq.~\eqref{eq: model 1-1}. 
Figure ~\ref{fig: dynamics of R_1}(a) illustrates the time evolution of the order parameter $R_1$.
%for $\omega_m = \omega_0(=0)$.
For $K_1=0.1, K_{\rm 2b}=0$, and $D=0.0$ (black solid line), we observe that $R_1$ is almost constant, indicating that the two-cluster state is stable.
However, in the presence of noise, we observe qualitatively different behaviors.
For $K_1=0.1$ and $D=0.1$ (orange solid line), $R_1$ slowly decreases and abruptly vanishes. Thus, in the presence of noise, the two-cluster state is actually not stable but meta-stable with a long lifetime.
We refer to this phenomenon as the noise erosion of the synchronized state because of its slow process. 
%Instead, the fully desynchronized state seems to be stable.
The evolution of the phase distribution during the process of this process are shown in Fig.~\ref{fig: distribution}(a). 
We qualitatively obtain the same results for small or vanishing $K_1$ values.
In contrast, for $K_1=0.3, K_{\rm 2b}=0$ and $D=0.1$ (blue solid line), $R_1$ seems to approach a particular nonvanishing value.
For this parameter set, we also test the initial condition of the fully desynchronized state (dotted green line) and observe the evolution of the phase distributions [Fig.~\ref{fig: distribution}(b)], suggesting that the system approaches a particular two-cluster state independent of the initial condition. When the type-b interaction is present (i.e., $K_{2 \mathrm{b}} >0$) instead of the one-simplex interaction, similar results are obtained for the synchronized initial conditions, as shown in Fig. \ref{fig: dynamics of R_1}(b). However, for the desynchronized initial condition, $R_1$ is vanishingly small for all $t>0$, indicating that the desynchronized state is stable. This behavior is preserved for larger $K_{\rm 2b}$ values (the result is not shown herein). 

%This indicates the bistability of the synchronized and 
%In Fig. \ref{fig: dynamics of R_1}(b), we plot the time evolution of $R_1$ for $\omega_m = \omega_0 = 0$. 
%We observe that the dynamics shown in Fig. \ref{fig: dynamics of R_1}(b) shares some of features with those in Fig. \ref{fig: dynamics of R_1}(a), except for the two differences. 
%The first one, which is less significant in the argument, is that the lifetime of two-cluster state with the same coupling strengths (i.e. the common $K_{2{\mathrm a}}$ and $K_1 = K_{2{\mathrm b}}$) is shorter for the orange lines. 
%We understand this quite readily by comparing the one-simplex term and the type-b interaction term and taking into consideration $R_2 \le 1$. 
%The second one is about the dotted green lines. In Fig. \ref{fig: dynamics of R_1}(a), $R_1$ rapidly increases from $0$, and the green line approaches the blue line. 
%On the other hand, in Fig. \ref{fig: dynamics of R_1} (b), the green line hovers around $0$ from beginning to end: synchronization does not emerge.
%  This difference implies that desynchronization-synchronization transition never occurs without one-simplex interaction, 
%and we understand this phenomenon quite intuitively as below: because the type-b interaction is "nonlinear", which means the effective coupling strength contains the product of the order parameters, 
%this term does not affect the linear stability of the incoherent state $R_1 = 0$.

Next, we consider the case of nonidentical oscillators. 
Specifically, the natural frequencies $\omega_m$ are drawn from a Lorentzian distribution $g(\omega)$ with mean $\omega_0$ and width $\gamma$, i.e. $g(\omega) = \frac{\gamma}{\pi [(\omega-\omega_0)^2 + \gamma^2]}$.
As shown in Figs. \ref{fig: dynamics of R_1} (c, d), we observe qualitatively similar features. 
Therefore, we expect that the cases of identical oscillators capture the essential properties of the system, and henceforth, we focus on the simpler case $\omega_m=\omega_0$ for ease of analysis. 
% the model is invariant under the transformations $\theta_m \to \theta_m-\omega_0 t$, $ t \to c \gamma t$, $K_i \to \frac{K_i}{c \gamma}$, and $D \to \frac{D}{c \gamma}$, with $c$ being an arbitrary constant.

% Next, we consider the case of nonidentical oscillators. Specifically, the natural frequencies $\omega_m$ are drawn from a Lorentzian distribution. 
% As shown in \footnote{See Supplemental Material at [URL will be inserted by publisher] for details}, quantitatively similar features are observed. 
% Therefore, we expect that the essential properties of the system are captured by the cases of identical oscillators, and 
% henceforth, we focus on the simpler case $\omega_m=\omega_0$ for ease of analysis.
% We may also fix $D$ to an arbitrary value without loss of generality. Henceforth, we set $D=0.1$ in numerical analysis. 

% If the intrinsic frequencies are drawn from a Lorentzian distribution, similar dynamics are observed, as shown in Fig. \ref{sfig: Lorentzian}. Essentially, 
% $\omega_m \sim g(\omega)$, where $g(\omega) = \frac{\gamma}{\pi [(\omega-\omega_0)^2 + \gamma^2]}$, $\omega_0$ is the mean frequency, and $\gamma$ is the width of the Lorentzian distribution. 
% We may set arbitrary $\omega_0$ and $\gamma$ values without loss of generality because 
% the model is invariant under the transformations $\theta_m \to \theta_m-\omega_0 t$, $ t \to c \gamma t$, $K_i \to \frac{K_i}{c \gamma}$, and $D \to \frac{D}{c \gamma}$, with $c$ being an arbitrary constant.
% Specifically, we set $\omega_0=0$ and $\gamma=0.01$.

\section{Analysis for stationary state}
\subsection{Stationary distribution}
We now establish a theory for understanding the synchronization and desynchronization processes.
We consider a continuum limit $N \to \infty$ for analytical tractability. Specifically, the number of oscillators with a phase within $(\theta, \theta+d\theta)$
is described by $N P(\theta, t) d\theta$, where $P(\theta, t)$ is a probability density function.
We start with the self-consistency approach to identify the steady states of the system. The order parameters are redefined as
\begin{align}
  \label{eq: order parameter continuous}
  % Z_l(t) = \int_{- \pi}^{\pi} \exp(il\theta) P(\theta, t) \mathrm{d} \theta\ \mathrm{for}\ l = 1, 2.
  Z_l(t) = \int_{- \pi}^{\pi} \exp(il\theta) P(\theta, t) \mathrm{d} \theta.
\end{align}
The Fokker--Planck equation equivalent to the Langevin equation \eqref{eq: model 1-1} is
\begin{align}
  \cfrac{\partial P}{\partial t} &= \cfrac{\partial}{\partial \theta}
  \left[ K_1 R_1 \sin (\theta-\Theta_1) + K_{2 \mathrm{a}} R_1^2\sin (2\theta-2\Theta_1) + K_{2 \mathrm{b}} R_1 R_2 \sin(\theta + \Theta_1 - \Theta_2)\right] P
  + D \cfrac{\partial^2 P}{\partial \theta^2}. \label{eq: FPE}
\end{align}
%From the results of simulations and analysis detailed in [appendix], we observe the simple relation between the two arguments $\Theta_1$ and $\Theta_2$, which is $\Theta_2 = 2 \Theta_1$.
%In addition, $\Theta_1(t)$ is a constant in the limit $N \to \infty$.  Thus, without loss of generality, we set $\Theta_1=\Theta_2=0$. Note that $Z_1(t)=R_1(t)$ for $\Theta_1=0$.
%As detailed in \cite{Note1}, we may assume $\Theta_2(t)=2\Theta_1(t)$. 
%Further, we numerically observe that $\Theta_1(t)$ evolves very slowly for large $N$\cite{Note1}, suggesting that $\Theta_1(t)$ is constant in the limit $N \to \infty$. If this is the case, we may assume $\Theta_1=\Theta_2=0$ without loss of generality. Note that $Z_l(t)=R_l(t)$ for $\Theta_l=0$.

As shown in Fig.~S2 in Appendix \ref{sec: Appendix A}, we numerically observe that $\Theta_1$ and $\Theta_2$ evolve slowly, suggesting that they are constant in the limit $N \to \infty$. 
Moreover, $\Theta_2=2 \Theta_1$ holds true for $t \ge 0$. 
We thus assume $\Theta_1=\Theta_2=0$.
Substituting
$\Theta_1=\Theta_2=0$ into Eq.~\eqref{eq: FPE}, we obtain
% the Fokker-Planck equation as follows:
\begin{equation}
  \label{seq: FPE reduced}
  \cfrac{\partial P(\theta, t)}{\partial t} = \cfrac{\partial}{\partial \theta} [\{K_1 R_1 \sin \theta + K_{2 \mathrm{a}} R_1^2 \sin 2 \theta + K_{2 \mathrm{b}} R_1 R_2 \sin \theta\}P] + D \cfrac{\partial^2 P}{\partial \theta^2}.
\end{equation}
The stationary distribution $P_{\mathrm{s}}(\theta)$ is found as a solution to $\partial_t P = 0$. The general solution is
%\begin{align}
%  \cfrac{\partial}{\partial \theta} \{ (K_1 R_1 \sin \theta + K_{2 \mathrm{a}} R_1^2 \sin 2 \theta)P \} + D \cfrac{\partial^2 P}{\partial \theta^2} = 0.
%  \label{seq: stationary 1}
%\end{align}
%The solution to Eq. \eqref{seq: stationary 1} is
\begin{align}
  P_{\mathrm{s}}(\theta) = &\exp \left(\cfrac{2 K_1 R_1 \cos \theta + K_{2 \mathrm{a}} R_1^2 \cos 2 \theta  + 2 K_{2 \mathrm{b}} R_1 R_2 \sin \theta}{2 D}\right) \cdot \notag\\
  &\left[ c_1 + c_2 \int_{-\pi}^{\theta} \exp \left(-\cfrac{2 K_1 R_1 \cos y + K_{2 \mathrm{a}} R_1^2 \cos 2 y  + 2 K_{2 \mathrm{b}} R_1 R_2 \sin y}{2 D}\right) \mathrm{d}y \right],
\end{align}
where $c_1$ and $c_2$ are constants.
Because $P_{\mathrm{s}} (\theta) = P_{\mathrm{s}} (\theta + 2 \pi)$ for any $\theta$, $c_2$ vanishes.
%Therefore, we obtain
%\begin{align}
%  \label{seq: stationary distribution appendix}
%  P_{\mathrm{s}}(\theta) = c_1 \exp\left( \cfrac{2 K_1 R_1 \cos \theta + K_{2 \mathrm{a}} R_1^2 \cos 2 \theta}{2D} \right),
%\end{align}
%where $c_1$ is obtained by the normalizing constant:
Thus, we obtain the stationary distribution $P_{\mathrm{s}}$ as 
\begin{align}
  \label{eq: stationary distribution}
  P_{\mathrm{s}}(\theta) = c_1 \exp\left( \cfrac{2 K_1 R_{1\mathrm{s}} \cos \theta + K_{2 \mathrm{a}} R_{1\mathrm{s}}^2 \cos 2 \theta + 2 K_{2 \mathrm{b}} R_{1\mathrm{s}} R_{2\mathrm{s}} \cos \theta}{2D} \right),
\end{align}
where $c_1$ is the normalizing constant given as
\begin{align}
  c_1 = \cfrac{1}{\displaystyle\int_{- \pi}^{\pi} \exp \left(\cfrac{2 K_1 R_1 \cos y + K_{2 \mathrm{a}} R_1^2 \cos 2y + 2 K_{2 \mathrm{b}} R_1 R_2 \cos y}{2 D}\right) \mathrm{d}y}.
  \label{seq: constant 2}
\end{align}

% Under this assumption, we have $Z_l(t)=R_l(t)$ for $\Theta_l=0$, and the stationary distribution $P=P_{\mathrm{s}}(\theta)$ for constants
% $R_1=R_{1\mathrm{s}}$ and $R_2 = R_{2 \mathrm{s}}$
% can be obtained by setting $\partial_t P = 0$, yielding
% \begin{align}
%   \label{eq: stationary distribution}
%   P_{\mathrm{s}}(\theta) = c \exp\left( \cfrac{2 K_1 R_{1\mathrm{s}} \cos \theta + K_{2 \mathrm{a}} R_{1\mathrm{s}}^2 \cos 2 \theta + 2 K_{2 \mathrm{b}} R_{1\mathrm{s}} R_{2\mathrm{s}} \cos \theta}{2D} \right),
% \end{align}
% where $c$ is a normalizing constant \cite{Note1}.
Substituting Eq. \eqref{eq: stationary distribution} into Eq.~\eqref{eq: order parameter continuous}, 
we obtain a set of self-consistent equations for
%$R_{1\mathrm{s}}$ and $R_{2\mathrm{s}}$, given as
$R_{l \mathrm{s}}$ ($l=1,2$), given as
\begin{align}
 \label{eq: selfconsistency 1}
  R_{l\mathrm{s}} &= %\int_{-\pi}^{\pi} P_{\rm s}(\theta; R_1, R_2) \cos l\theta\ {\rm d} \theta =: S_l(R_{1\mathrm{s}}, R_{2\mathrm{s}}).
\int_{-\pi}^{\pi} P_{\rm s}(\theta) \cos l\theta\ {\rm d} \theta =: S_l(R_{1\mathrm{s}}, R_{2\mathrm{s}}).
%  \label{eq: selfconsistency 2}
%  R_{2\mathrm{s}} &= \int_{-\pi}^{\pi} P_{\rm s}(\theta; R_1, R_2) \cos 2\theta\ {\rm d} \theta =: S_2(R_{1\mathrm{s}}, R_{2\mathrm{s}}),
\end{align}
%We note that the set of self-consistent equations (\ref{eq: selfconsistency 1},\ref{eq: selfconsistency 2}) is reduced to the one equation for $R_{1 \mathrm{s}}$ (i.e. Eq.\eqref{eq: selfconsistency 1}) when $K_{2 \mathrm{b}} = 0$.
We observe that the right-hand side of Eq.~\eqref{eq: selfconsistency 1} for $l=1$ vanishes for 
$K_1=0$ and $K_{2 \mathrm{b}}=0$ because then $P_{\mathrm{s}}(\theta)$ is $\pi$-periodic.
This implies that for $K_1=0$ and $K_{2 \mathrm{b}}=0$, there is no stationary distribution except that corresponding to $R_{1\mathrm{s}}=0$, which is the uniform distribution $P_{\rm s}(\theta)=\frac{1}{2\pi}$ \cite{Komarov2015-es}.
In this case, $R_2$ also vanishes.
Nontrivial steady distributions may only arise when $K_1>0$ or $K_{2 \mathrm{b}} > 0$.

\subsection{Bifurcation analysis}
\label{subsec: Bifurcation}
We investigate the bifurcation of the system by numerically solving Eq.~\eqref{eq: selfconsistency 1}.
%for $R_{1\mathrm{s}}$ in the case of $K_{2 \mathrm{b}} = 0$.
First, we analyze the case of $K_{2 \mathrm{b}} = 0$. Here, we only need to consider Eq.~\eqref{eq: selfconsistency 1} for $l=1$ and numerically identifying $R_{1\mathrm{s}}$ is straightforward (see Appendix \ref{sec: Appendix B}). 
In Fig.~\ref{fig:phase diagram}(a), the phase diagram in the $(K_1,K_{2 \mathrm{a}})$ plane with $K_{2 \mathrm{b}} = 0$ is displayed, suggesting that a bifurcation occurs at 
$K_1=K_{\rm c}=0.2$ for all $K_{2 \mathrm{a}}$. 
In Figs.~\ref{fig:phase diagram}(c) and (d), we plot $R_{1\mathrm{s}}$ as a function of $K_1$ for $K_{2 \mathrm{a}}=0.15$ and $K_{2 \mathrm{a}}=3.0$, respectively. 
We observe that
supercritical and subcritical pitchfork bifurcations occur in Figs.~\ref{fig:phase diagram}(c) and (d), respectively. Moreover, the bifurcation type appears to change from supercritical to subcritical at $K_1=K_{\rm c}$, yielding the bistable region where both desynchronized and synchronized states are stable for $K_1>K_{\rm c}$. Note that $K_{\rm c}=0.2$ is equal to $2D$. 
Next, we consider $K_{2 \mathrm{b}} > 0$. By numerically solving Eq.~\eqref{eq: selfconsistency 1} for $l=1,2$ (See Appendix \ref{sec: Appendix B}), we obtain the $R_{1\mathrm{s}}$ value of the synchronized state for given $K_{2 \mathrm{a}}$ and $K_{2 \mathrm{b}}$ values, which is depicted in Fig.~\ref{fig:phase diagram}(b). 
\begin{figure}[htb]
  \begin{tabular}{cc}
    \begin{minipage}[t]{5cm}
      \centering
      \includegraphics[height = 4cm]{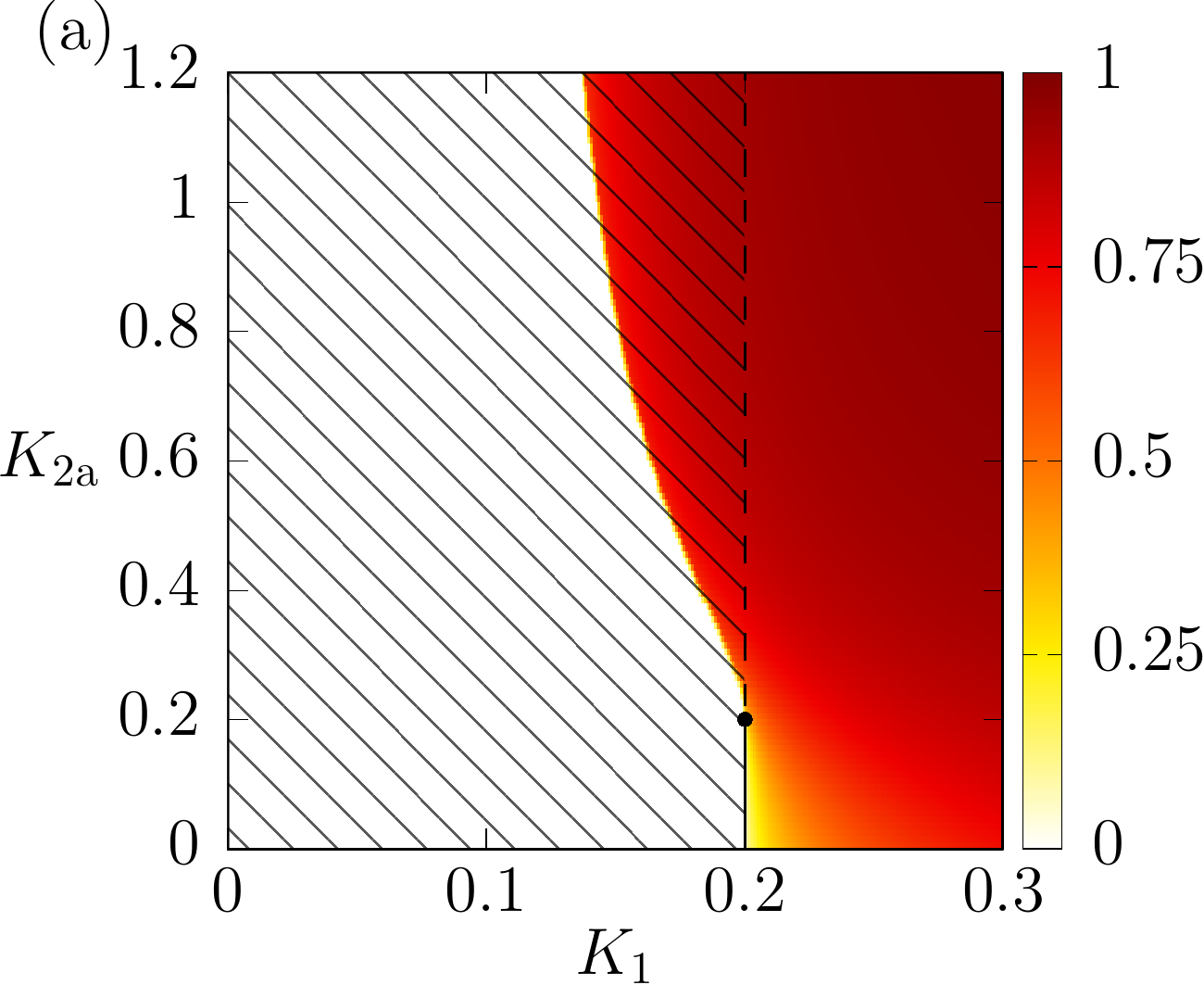}
    \end{minipage} &
    \hspace{1cm}
    \begin{minipage}[t]{5cm}
      \centering
      \includegraphics[height = 4cm]{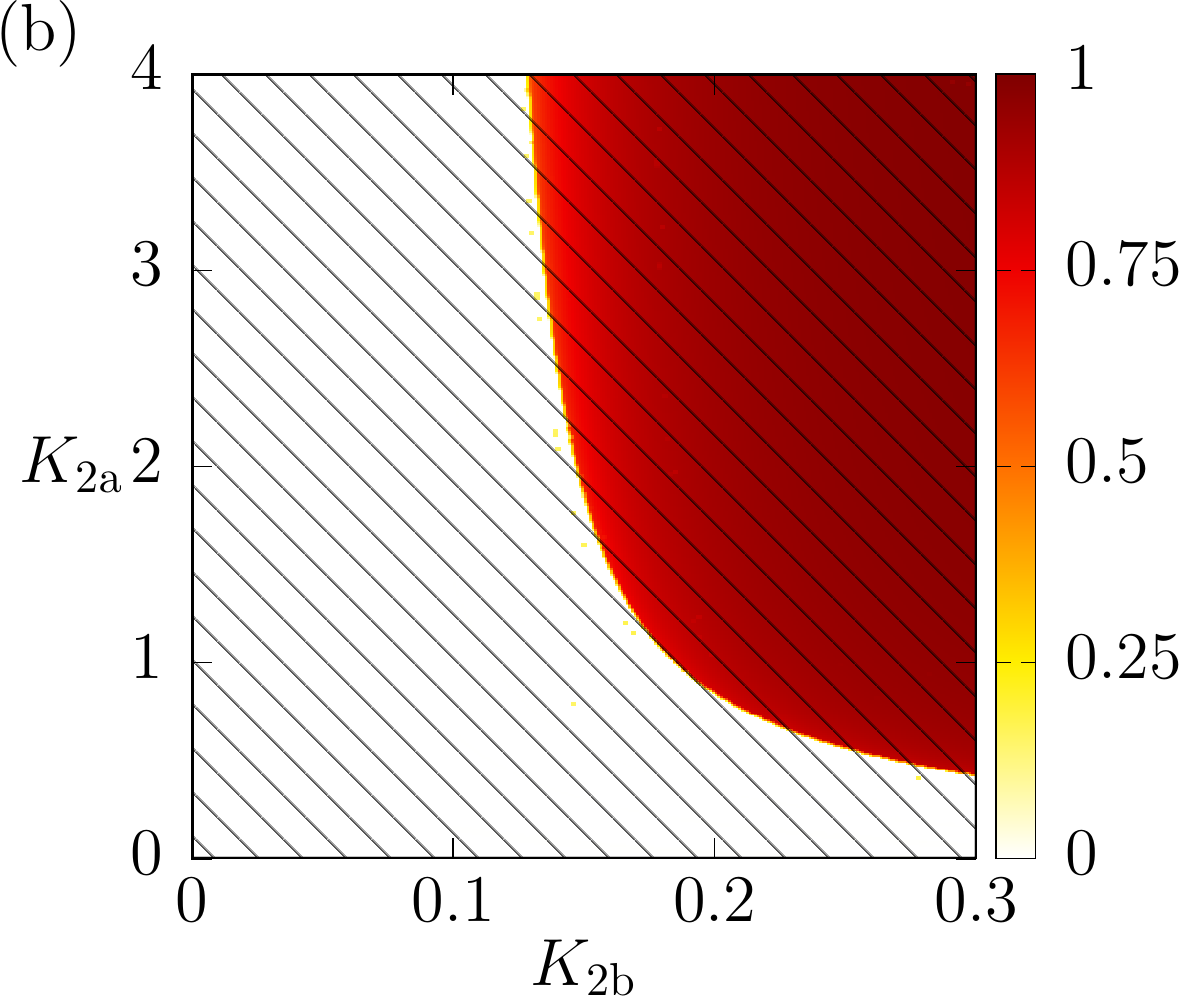}
    \end{minipage}\\
    \begin{minipage}[t]{5cm}
      \centering
      \includegraphics[height = 4cm]{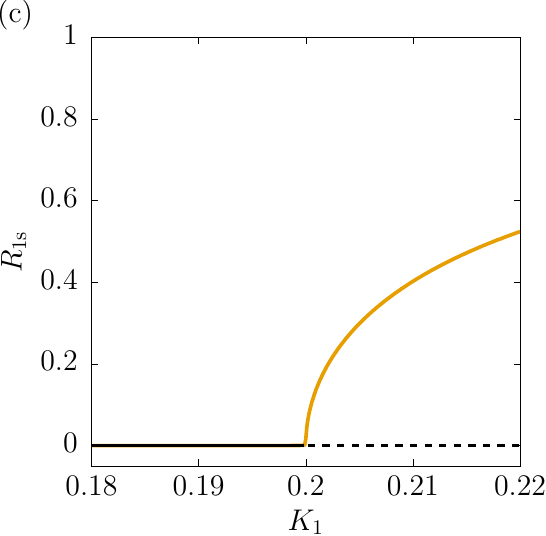}
    \end{minipage}&
    \hspace{1cm}
    \begin{minipage}[t]{5cm}
      \centering
      \includegraphics[height = 4cm]{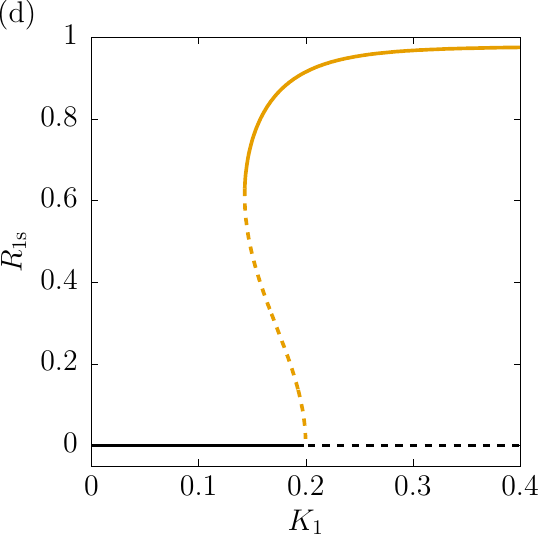}
    \end{minipage}
  \end{tabular}
  \caption{%(a) Typical behavior of the self-consistency equation \eqref{eq: selfconsistency 1} for $K_{2 \mathrm{a}} = 1.0$ with varying $K_1$.
  (a,b) Phase diagrams in (a) the $(K_1, K_{2 \mathrm{a}})$ plane for $K_{2 \mathrm{b}} = 0$ and (b) the $(K_{2 \mathrm{b}}, K_{2 \mathrm{a}})$ plane for $K_1 = 0$. The color scale describes the $R_1$ value of the stable synchronized state.
The hatched region denotes the region where the desynchronized state is stable. 
 In (a), the vertical line at $K_{2 \mathrm{a}}=0.2$ denotes the critical coupling strength above which the desynchronized state is unstable. The solid and dashed lines denote the supercritical and subcritical bifurcation curves, respectively.
The hatched colored region denotes the bistable region where both synchronized and desynchronized states are stable, and the dot denotes the bifurcation type switch at $K_1 = 2 D$, as described by Eq.~\eqref{eq: bifurcation}.
%  (b) Phase diagram in the $(K_{2 \mathrm{b}, K_{2 \mathrm{a}}})$ plane for $K_1 = 0$.
%The color scale describes the $R_1$ value of the stable synchronized state.
(c, d) Bifurcation diagrams for (c) $K_{2 \mathrm{a}} = 0.15$ and (d) $K_{2 \mathrm{a}} = 1.0$, where we fix $K_{2 \mathrm{b}} = 0$. In all the panels, we fix $D=0.1$.
  }
  \label{fig:phase diagram}
\end{figure}

% To elucidate the bifurcation structure,
% % when $K_{2 \mathrm{b}} = 0$,
% we perform a weakly nonlinear analysis using a standard method as follows \cite{Kuramoto1975-qj,Pikovsky2001-hv}. 
% %We denote the bifurcation point of $R_1=Z_1=0$ by $K_1=K_{\rm c}$ and set $K_1 = K_{\rm c} (1+\mu)$, where $\mu$ is the bifurcation parameter.
% % By substituting Fourier modes of the density $P(t)$ into Eq. \eqref{eq: FPE}, we obtain an equation for the complex order parameter $Z_1$ as
% % We first expand the distribution $P(\theta, t)$ in series with respect to the order parameter $Z_l$ as
% This analysis is carried out around the incoherent solution $Z_l = R_l = 0$.
% We do not set the intrinsic frequency in the analysis below as in earlier analyses. This approach ensures that the methodologies remain applicable to more general models with additional coupling terms. 

To elucidate the bifurcation structure, we perform a weakly nonlinear analysis by applying a previously proposed methodology \cite{Kuramoto2003-xb} to our model. 
This analysis is carried out around the incoherent solution $Z_l = R_l = 0$ and is valid near the bifurcation point at which the solution becomes unstable. 
We will introduce a bifurcation parameter, which is treated as the sole small parameter in our analysis. 
Although the value of $\omega$ may arbitrarily be chosen and was set to $0$ because of the translational symmetry in our model, we here assume $\omega = O(1)$ to apply the methodology proposed in \cite{Kuramoto2003-xb}. 
This methodology can even apply to models where coupling terms break the translational symmetry, an advantage for future extensions.

Specifically, we expand the distribution $P(\theta, t)$ in series with respect to the order parameter $Z_l$ as
\begin{align}
  \label{seq: Fourier series}
  P(\theta, t) = \cfrac{1}{2 \pi} \left(1 + \sum_{l \ne 0} Z_{-l} (t) e^{il \theta} \right),
\end{align}
where $Z_{-l}= \bar Z_l$. This expansion is analogous to a Fourier series expansion where $Z_l$ corresponds to the Fourier coefficients.
% By substituting Eq. \eqref{seq: Fourier series} into Eq. \eqref{eq: FPE}, we obtain an equation for the complex order parameter $Z_1$ as
% \begin{align}
%   \label{eq: Fourier eq}
%   % \dot{Z}_1 = \cfrac{K_1 - 2D}{2} Z_1 - \cfrac{K_1 P_{-2} P_1}{2} - \cfrac{K_{2 \mathrm{a}}}{2} (P_1 Z_1^2 - P_{-3} P_1^2)
%   %  - \cfrac{K_{2 \mathrm{b}}}{2}(P_2 P_{-1} P_{-2} - P_{-2} P_1),
%   \dot{Z}_1 = \cfrac{K_1 - 2D}{2} Z_1 - \cfrac{K_1 Z_2 Z_{-1}}{2} - \cfrac{K_{2 \mathrm{a}}}{2} (Z_3 Z_{-1}^2 - Z_{-1} Z_1^2) - \cfrac{K_{2 \mathrm{b}}}{2}(Z_1 Z_{-2} Z_2 - Z_{-1} Z_2).
% \end{align}
%where $P_{l}$ is a Fourier mode and defined as
% where $P_{l}$ is the Fourier mode defined as
% \begin{align}
%   \label{eq: Fourier mode}
%   P_l(t) = \int_{-\pi}^{\pi} P(\theta, t) \exp(i l \theta) \mathrm{d} \theta.
% \end{align}
% Note that $Z_1 = P_{-1}$. 
Using the complex order parameters for $l = \pm 1, \pm 2$, the Fokker-Plank equation \eqref{seq: FPE reduced} can be rewritten as
\begin{align}
  \label{seq: FPE 2}
  \cfrac{\partial P}{\partial t} = \cfrac{1}{2i} 
  \cfrac{\partial}{\partial \theta} \big(\{& -2 i \omega + K_1 [Z_{-1} \exp(i \theta) - Z_1 \exp(-i\theta)] 
   + K_{2 \mathrm{a}} [Z_{-1}^2 \exp(2 i \theta) - Z_1^2 \exp(-2i\theta)]  \notag \\
   + &K_{2 \mathrm{b}} [Z_{-2} Z_1 \exp(i \theta) - Z_{-1} Z_2 \exp(-i \theta)]
   \}P \big) + D \cfrac{\partial^2 P}{\partial \theta^2}.
\end{align}
As mentioned, the term involving $\omega$ remains in Eq. \eqref{seq: FPE 2}.
By further substituting Eq.~\eqref{seq: Fourier series} into Eq.~\eqref{seq: FPE 2}, we obtain
\begin{align}
  \cfrac{\mathrm{d} Z_1}{\mathrm{d} t} = &(- i \omega - D)Z_1 - \cfrac{K_1}{2} (Z_2 Z_{-1} - Z_1)  %\notag \\
  - \cfrac{K_{2 \mathrm{a}}}{2} (Z_3 Z_{-1}^2 - Z_{-1} Z_1^2) - \cfrac{K_{2 \mathrm{b}}}{2}(Z_1 Z_{-2} Z_2 - Z_{-1} Z_2), \notag\\
  \label{seq: FPE Fourier l=1}
  = &\cfrac{K_1 - 2D - 2 i \omega}{2} Z_1 - \cfrac{K_1 Z_2 Z_{-1}}{2} - \cfrac{K_{2 \mathrm{a}}}{2} (Z_3 Z_{-1}^2 - Z_{-1} Z_1^2) - \cfrac{K_{2 \mathrm{b}}}{2}(Z_1 Z_{-2} Z_2 - Z_{-1} Z_2)\\
  \label{seq: FPE Fourier l=2}
  \cfrac{\mathrm{d} Z_2}{\mathrm{d} t} = &(- 2 i \omega - 4 D) Z_2 - K_1 (Z_3 Z_{-1} - Z_1^2) %\\
  - K_{2 \mathrm{a}} (Z_4 Z_{-1}^2 - Z_1^2) - K_{2 \mathrm{b}} (Z_1 Z_{-2} Z_3 - Z_{-1} Z_2 Z_1),\\
  \label{seq: FPE Fourier l}
  \cfrac{\mathrm{d} Z_l}{\mathrm{d} t} = &(- i l \omega - l^2 D) Z_l - \cfrac{l K_1}{2} (Z_{l+1} Z_{-1} - Z_{l-1} Z_1) \notag \\
  &- \cfrac{l K_{2 \mathrm{a}}}{2}(Z_{l+2} Z_{-1}^2 - Z_{l-2} Z_1^2) - \cfrac{l K_{2 \mathrm{b}}}{2}(Z_1 Z_{-2} Z_{l+1} - Z_{-1} Z_2 Z_{l-1})\ \  \mbox{for $l \ne \pm 1, \pm 2$}.
\end{align}
Note that $Z_{-1}$ and $Z_{-2}$ obey the complex conjugate of the right hand side of Eqs.~\eqref{seq: FPE Fourier l=1} and Eq.~\eqref{seq: FPE Fourier l=2}, respectively.

Equation \eqref{seq: FPE Fourier l=1} implies that the state
%$Z_1=0$
$Z_1=0$
bifurcates at $K_1=2D$; thus we set
%Let $K_1$ a control parameter and $\mu$ be a bifurcation parameter:
\begin{align}
  \label{seq: bifurcation parameter}
  K_1 = 2 D(1 + \mu),
\end{align}
where $\mu$ is the bifurcation parameter.
We introduce $\varepsilon = \sqrt{|\mu|}$ and the scaled time $\tau = \varepsilon^2 t$.
The time derivative then transformed as
\begin{align}
  \label{seq: time differentiation}
  \cfrac{\mathrm{d}}{\mathrm{d}t} \rightarrow \cfrac{\partial}{\partial t} + \varepsilon^2 \cfrac{\partial}{\partial \tau}.
\end{align}
We expand the order parameter $Z_l(t)$ into $Z_{l,\nu}(t, \tau)$ as
\begin{align}
  \label{seq: expansion}
  Z_{l}(t, \tau) = \varepsilon Z_{l,1} (t, \tau) + \varepsilon^2 Z_{l,2} (t, \tau) + \cdots.
\end{align}
% Order parameter $Z_1$ is also expanded as
% \begin{align}
%   \label{seq: order parameter expansion}
%   Z_1(t, \tau) = \varepsilon Z_{1, 1} (t, \tau) + \varepsilon^2 Z_{1, 2} (t, \tau) + \cdots.
% \end{align}
Substituting Eqs. \eqref{seq: bifurcation parameter}--\eqref{seq: expansion} into the series \eqref{seq: FPE Fourier l=2}--\eqref{seq: FPE Fourier l},
we obtain
%yields the system of balance equations: 
\begin{align}
  \label{seq: Z_l}
  \left( \cfrac{\partial}{\partial t} + i l \omega + l^2  D\right) Z_{l,\nu} &= B_{l, \nu},\ \ (l \ne \pm 1)\\
  \label{seq: Z_1}
  \left(\cfrac{\partial}{\partial t} \pm i \omega \right) Z_{\pm 1, \nu} &= B_{\pm 1, \nu},\ \ 
\end{align}
where
%$B_{l, \nu}$ has the forms except for $l = 1, 2$: 
\begin{align}
  \label{seq: B_1_1}
  B_{1,1} = &\ 0,\\
  \label{seq: B_1_2}
  B_{1,2} = &\left(- D+ \cfrac{K_{2 \mathrm{b}}}{2} \right) Z_{2,1} Z_{-1,1},\\
  \label{seq: B_1_3}
  B_{1,3} = & \left( - \cfrac{\partial}{\partial \tau} \pm
  D \right) Z_{1,1}
  - D(Z_{2,2} Z_{-1,1}
  + Z_{2,1} Z_{-1,2}) \notag\\
  &- \cfrac{K_{2 \mathrm{a}}}{2}(Z_{3,1} Z_{-1,1}^2 - Z_{-1,1} Z_{1,1}^2) 
  - \cfrac{K_{2 \mathrm{b}}}{2}(Z_{-2,1} Z_{1,1} Z_{2,1} - Z_{2,2} Z_{-1,1} - Z_{2,1} Z_{-1,2}),\\
  \label{seq: B_2_2}
  B_{2,2} = &2 D(Z_{1, 1}^2 - Z_{3,1} Z_{-1,1}) + K_{2 \mathrm{a}}  Z_{1, 1}^2,
\end{align}
and, for $l \neq \pm 1, \pm 2$,
\begin{align}
  \label{seq: B_l_1}
  B_{l, 1} =&\ 0,\\
  \label{seq: B_l_2}
  B_{l, 2} =&\ l D (Z_{l-1,1} Z_{1,1} - Z_{l+1,1} Z_{-1,1}),\\
  \label{seq: B_l_3}
  B_{l, 3} =& - \cfrac{\partial}{\partial \tau} Z_{l,1} 
  +l D (Z_{l-1,1} Z_{1,2} + Z_{l-1,2} Z_{1,1}
%\notag\\ &
  - Z_{l+1,1} Z_{-1,2} - Z_{l+1,2} Z_{-1,1}) 
  \notag\\
  &- \cfrac{l K_{2 \mathrm{a}}}{2} (Z_{l+2,1} Z_{-1,1}^2 - Z_{l-2,1} Z_{1,1}^2)
  - \cfrac{l K_{2 \mathrm{b}}}{2} (Z_{-2,1} Z_{1,1} Z_{l+1,1} - Z_{2,1} Z_{-1,1} Z_{l-1,1}).
\end{align}
Note that the plus-minus sign in Eq. \eqref{seq: B_1_3} corresponds to the sign of $\mu$.

Since $\exp(\mp i \omega t)$ is an eigenfunction of the operator $(\partial / \partial t \pm i \omega)$ on the left-hand side of Eq. \eqref{seq: Z_1} with an eigenvalue 0, the right-hand side has no corresponding component. 
We obtain the solvability condition from this fact as
\begin{align}
  \label{eq: solvability condition}
  \int_0^{2 \pi/\omega} B_{1, \nu}(t, \tau) \exp(i \omega t)  \mathrm{d} t = \int_0^{2 \pi/\omega} B_{-1, \nu}(t, \tau) \exp(- i \omega t)  \mathrm{d} t = 0.
\end{align}
If $B_{1, \nu}$ are expanded into Fourier series as
\begin{align}
  \label{eq: expansion of B}
  B_{\pm 1, \nu} = \sum_{m = - \infty}^{\infty} B_{\pm 1, \nu}^{(m)} \exp(i m \omega t), 
\end{align}
the solvability condition then reduces to 
\begin{align}
  \label{eq: solvability condition 2}
  B_{1, \nu}^{(-1)} = B_{-1, \nu}^{(1)} = 0.
\end{align}

We solve the system of Eqs. \eqref{seq: Z_l} and \eqref{seq: Z_1}.
% Since $Z_1(t, \tau)$ is a function of $\tau$ because of Eqs. (\ref{seq: Z_1}-\ref{seq: B_l_1}), 
% we rewrite $Z_1$ as $Z_1(\tau)$.
Because of Eqs. \eqref{seq: Z_1} and \eqref{seq: B_1_1}, %$Z_{1,1}$ are independent of $t$. 
\begin{align}
  \label{eq: Z_{1,1}}
  Z_{1,1} = \overline{Z_{-1, 1}} = W(\tau) \exp(- i \omega t),
\end{align}
where $W(\tau)$ has not yet been specified.
% Thus, we write $Z_{1,1}(t, \tau)$ as $Z_{1,1}(\tau)$.
Next, we obtain
\begin{align}
  \label{seq: Z_{3,1}}
  Z_{3,1} = 0
\end{align}
because of Eqs. \eqref{seq: Z_l} and \eqref{seq: B_l_1} for $l = 3$. By substituting Eq. \eqref{seq: Z_{3,1}} into Eq. \eqref{seq: B_2_2}, $B_{2,2}$ reduces to
\begin{align}
  B_{2,2} &= (2 D + K_{2 \mathrm{a}}) Z_{1, 1}^2 \\
  &= (2 D + K_{2 \mathrm{a}}) W(\tau)^2 \exp(- 2 i \omega t).
\end{align}
From Eq. \eqref{seq: Z_l}, we have the equation for $Z_{2,2}$ as
\begin{align}
  \label{seq: Z_{2,2}}
  \left( \cfrac{\partial}{\partial t} + 4  D\right) Z_{2,2} = (2 D + K_{2 \mathrm{a}}) W(\tau)^2 \exp(- 2 i \omega t),
\end{align}
and 
\begin{align}
  Z_{2,2} = \cfrac{(2 D + K_{2 \mathrm{a}}) W(\tau)^2 \exp(- 2 i \omega t)}{4 D}
\end{align}
is a long time solution to Eq. \eqref{seq: Z_{2,2}}.
In the same way as $Z_{3,1}$, we obtain
%$Z_{2,1}$ vanishes:
\begin{align}
  \label{seq: Z_{2,1}}
  Z_{2,1} = 0, %.
\end{align}
which leads to $B_{1, 2} = 0$. Thus, $Z_{1, 2} (t, \tau)$ is not dependent on $t$.
Substituting Eqs. \eqref{seq: Z_{3,1}}, \eqref{seq: Z_{2,2}} and \eqref{seq: Z_{2,1}}
into Eq.~\eqref{seq: B_1_3} yields 
\begin{align}
  \label{seq: B_{1,3}}
  B_{1,3}  
  %&= \left[ \left( - \cfrac{\partial}{\partial \tau} \pm 
  % D \right) W(\tau)
  % - \cfrac{(2 D + K_{2 \mathrm{a}}) W(\tau)^2 \overline{W(\tau)}}{4} + \cfrac{K_{2 \mathrm{a}} W(\tau)^2 \overline{W(\tau)}}{2} + \cfrac{K_{2 \mathrm{b}}(2D + K_{2 \mathrm{a}}) W(\tau)^2 \overline{W(\tau)}}{8D}\right] \exp(i \omega t)\\
  &= \left[ \left( - \cfrac{\partial}{\partial \tau} \pm
  D \right) W(\tau)
  - \cfrac{2D (2D - K_{2 \mathrm{b}}) - K_{2 \mathrm{a}}(2D + K_{2 \mathrm{b}})}{8D} |W(\tau)|^2 W(\tau) \right] \exp (- i \omega t) \\
  &= B_{1, 3}^{(-1)} \exp(- i \omega t).
\end{align}

From the solvability condition \eqref{eq: solvability condition 2} for $\nu = 3$, we obtain the normalized equation in the lowest order:
\begin{align}
  \label{seq: Z_1 2}
  \cfrac{\partial W(\tau)}{\partial \tau} = \pm D W(\tau) - \cfrac{2D (2D - K_{2 \mathrm{b}}) - K_{2 \mathrm{a}}(2D + K_{2 \mathrm{b}})}{8D} |W(\tau)|^2 W(\tau).
\end{align}
Finally, we obtain the equation with respect to $t$ in the lowest order, which is the normal form for the Hopf bifurcation as below:
\begin{align}
  \label{eq: amplitude eq}
  \cfrac{\mathrm{d} Z_1}{\mathrm{d}t} &= \cfrac{\partial Z_1}{\partial t} + \varepsilon^2 \cfrac{\partial Z_1}{\partial \tau} \\
  &= \varepsilon \cfrac{\partial Z_{1,1}}{\partial t} + \varepsilon^2 \cfrac{\partial Z_{1,2}}{\partial t} + \varepsilon^3 \cfrac{\partial Z_{1,3}}{\partial t} + + \varepsilon^3 \cfrac{\partial Z_{1,1}}{\partial \tau} + O(\varepsilon^4)\\
  &= \varepsilon B_{1,1} + \varepsilon^2 B_{1, 2} + \varepsilon^3 B_{1, 3} + \varepsilon^3 \cfrac{\partial Z_{1, 1}}{\partial \tau} + O(\varepsilon^4)\\
  &=  \varepsilon^3 \cfrac{\partial Z_{1, 1}}{\partial \tau} + O(\varepsilon^4)\\
  &= \varepsilon^3 \cfrac{\partial W(\tau)}{\partial \tau} \exp(- i \omega t) + O(\varepsilon^4)\\
  &= \cfrac{K_1 - 2 D}{2} Z_1 - g |Z_1|^2 Z_1 + O(\varepsilon^4),
\end{align}
where
\begin{align}
  g = \cfrac{2D (2D - K_{2 \mathrm{b}}) - K_{2 \mathrm{a}}(2D + K_{2 \mathrm{b}})}{8D}.
\end{align}

% According to Eq. \eqref{eq: Fourier eq}, we find that the desynchronized state $Z_1 = 0$ is destabilized at $K_1 = K_{\rm c}$, where $K_{\rm c} = 2D$.
%We set $K_1 = K_{\rm c} (1+\mu)$, where $\mu$ is the bifurcation parameter.
%The complex order parameter $Z_1$ is expanded as
%\begin{align}
% %Z_1 = \varepsilon Z_{1, 1} + \varepsilon^2 Z_{2, 1} + \cdots,
% Z_1 = \varepsilon Z_{1, 1} + \varepsilon^2 Z_{1, 2} + \cdots,
%\end{align}
%where $\varepsilon=\sqrt{|\mu|}$.
%As detailed in \cite{Note1}, we derive
%\begin{align}
%\label{eq: amplitude eq}
% \cfrac{1}{\varepsilon^2} \dot Z_{1, 1} 
%= \cfrac{K_1-2 D}{2}\ Z_{1, 1} - g|Z_{1, 1}|^2 Z_{1, 1},
%\end{align}
% As detailed in \cite{Note1}, we derive
% \begin{align}
% \label{eq: amplitude eq}
%  \dot R_{1} 
% = \cfrac{K_1-2 D}{2}\ R_{1} - g|R_{1}|^2 R_{1},
% \end{align}
% where
% \begin{align}
%   %g &= \cfrac{K_{\rm c}^2 + K_{\rm c} K_{2 \mathrm{a}}}{8D} - \cfrac{K_{2 \mathrm{a}}}{2}.
%  %g &= 2 K_{2 \mathrm{a}}(1+2 K_{2 \mathrm{b}}) - 4D(1-K_{2 \mathrm{b}}).
%  g &=  \cfrac{2D (2D - K_{2 \mathrm{b}}) - K_{2 \mathrm{a}}(2D + K_{2 \mathrm{b}})}{8D}.
% \end{align}
Note that $g$ is real in this particular system.
The sign of $g$ determines the bifurcation type; supercritical and subcritical bifurcations occur for $g>0$ (or $K_{2 \mathrm{a}}<K_{2 \mathrm{a}}^*$) and $g<0$ (or $K_{2 \mathrm{a}}>K_{2 \mathrm{a}}^*$), respectively, where
\begin{gather}
  \label{eq: bifurcation}
 K_{2 \mathrm{a}}^* = \frac{2D (2D-K_{2 \mathrm{b}})}{2D+K_{2 \mathrm{b}}},
\end{gather}
which assumes $2D$ for $K_{2 \mathrm{b}}=0$ and decreases with increasing $K_{2 \mathrm{b}}$.
%This theoretical analysis is in perfect agreement with the numerical analysis of Eq.~\eqref{eq: selfconsistency 1} shown in Fig.~\ref{fig:phase diagram}.
This theoretical analysis clarifies the parameter region of the stable desynchronized state and the change in the bifurcation nature at $K_{2 \mathrm{a}}=K_{2 \mathrm{a}}^*$, which are in perfect agreement with the phase diagrams shown in Figs.~\ref{fig:phase diagram} and \ref{sfig: phase diagram} in Appendix \ref{sec: Appendix C}.

\section{Transient dynamics: Time evolution of $R_1$}

We shift our focus to transient phenomena. 
As we observed in Fig. \ref{fig: dynamics of R_1}, the synchronized state slowly decays to the asynchroous state.
To elucidate the mechanism of this slow process, 
Eq.~\eqref{eq: amplitude eq} is not adequate because it is valid only around $R_1=0$.
We will demonstrate that under some assumptions including $K_{2 \mathrm{b}} = 0$,
a dynamical equation for $R_1$, which is 
approximately valid for large $R_1$,
may be obtained in a closed form, whereby we may determine the lifetime of the synchronized state in cases where the system transitions to the desynchronized state.
% Because the analysis for $K_1>0$, which is detailed in Appendix \ref{sec: Appendix D}, is complicated, 
% we here present the result for $K_1=0$.

\begin{figure}
  \centering
  \includegraphics[height = 5cm]{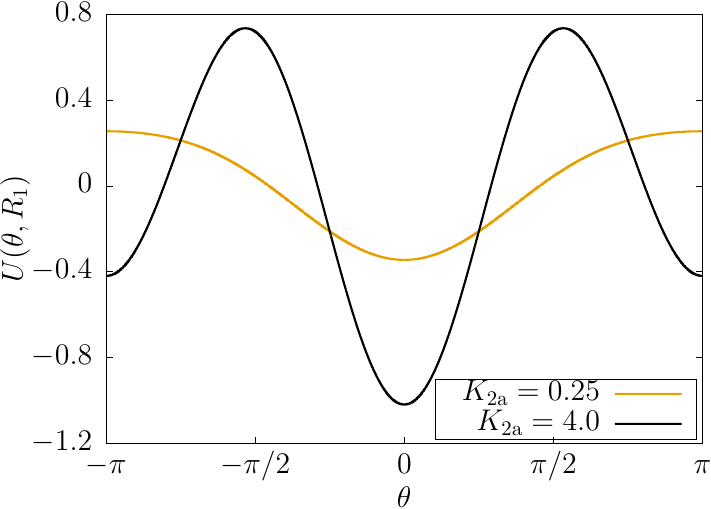}
  \caption{Graphical reprensentation of the potential $U(\theta, R_1)$. $R_1 = 0.6, K_1 = 0.50$.}
  \label{sfig: potential}
\end{figure}

% In the main text, we show the evolution equation of $R_1$ only for $K_1=0$ and $K_{2 \mathrm{b}} = 0$.
% Here, we derive this for $K_1\geq 0$.
We rewrite Eq.~\eqref{eq: model 1-1} as the following graident system:
\begin{equation}
 \dot \theta_m = -\cfrac{\partial}{\partial \theta_m} U(\theta_m,R_1) + \xi_m,
\end{equation}
where 
%The potential $U$ corresponding to Eq.~\eqref{seq: model 1-1} is given as
\begin{align}
  \label{seq: potential 2}
  U(\theta, R_1) = -\cfrac{1}{2} (K_{2 \mathrm{a}} R_1^2 \cos 2\theta + 2 K_1 R_1 \cos \theta).
%  \dot{\theta} &= \cfrac{\partial}{\partial \theta} U(\theta, R_1) + \xi_m.
\end{align}
We show typical $U$ shapes as a function of $\theta$ in Fig. \ref{sfig: potential}.
We focus on the case in which this function has two minima; i.e., $U$ is double-well.
Because 
\begin{align}
    \cfrac{\partial}{\partial \theta }U(\theta, R_1) &= K_{2 \mathrm{a}} R_1^2 \sin 2\theta + K_1 R_1 \sin \theta \notag\\
    % &= 2 K_{2 \mathrm{a}} R_1^2 \sin \theta \cos \theta + K_1 R_1 \sin \theta \notag\\
    &= R_1 \sin \theta (2 K_{2 \mathrm{a}} R_1 \cos \theta + K_1),
    \label{seq: derivative condition}
\end{align}
the necessary and sufficient condition for the potential to be double-well is
\begin{align}
  \label{seq: necessary condition}
  \cfrac{K_1}{2 K_{2 \mathrm{a}} R_1} < 1.
\end{align}
%graphical view of potentials for two cases: single-well and double-well in Fig. \ref{fig: potential}.
%In the analysis described
Below, we assume Eq.~\eqref{seq: necessary condition}.
The following quantities will be needed later.
The minima of the potential $U(\theta, R_1)$ are denoted as
\begin{align}
  U_{\min 1} &:= U(0, R_1) = - \cfrac{1}{2}(K_{2 \mathrm{a}} R_1^2 + 2 K_1 R_1),\\
  U_{\min 2} &:= U(\pi, R_1) = - \cfrac{1}{2}(K_{2 \mathrm{a}} R_1^2 - 2 K_1 R_1).
\end{align}
The maximum value $U_{\max}$ is
\begin{align}
  U_{\max} &:= U(\theta_{\max}, R_1) \notag\\
  &=-\cfrac{1}{2}\ (K_{2 \mathrm{a}} R_1^2 \cos 2 \theta_{\max} + 2 K_1 R_1 \cos \theta_{\max}) \notag\\
  &= \cfrac{1}{2} \left(K_{2 \mathrm{a}} R_1^2 + \cfrac{K_1^2}{2 K_{2 \mathrm{a}}}\right),
\end{align}
where $\theta_{\mathrm{max}}$ is defined as one of two maximum points of the potential $U(\theta)$ within the range $0<\theta<\pi$.
% We define $\theta_{\max}$ as one of two maximum points of the potential $U(\theta)$ within the range $0<\theta<\pi$.
The potential barriers are
\begin{align}
  \Delta U_1 &:= U_{\max} - U_{\min 1} \notag \\
% &= \cfrac{1}{2} \left(K_{2 \mathrm{a}} R_1^2 + \cfrac{K_1^2}{2 K_{2 \mathrm{a}}}\right) + \cfrac{1}{2} (K_{2 \mathrm{a}} R_1^2 + 2 K_1 R_1) \notag\\
  &= K_{2 \mathrm{a}} R_1^2 + K_1 R_1 + \cfrac{K_1^2}{4 K_{2 \mathrm{a}}},\\
  \Delta U_2  &:= U_{\max} - U_{\min 2} \notag \\
% &= \cfrac{1}{2} \left(K_{2 \mathrm{a}} R_1^2 + \cfrac{K_1^2}{2 K_{2 \mathrm{a}}}\right) + \cfrac{1}{2} (K_{2 \mathrm{a}} R_1^2 - 2 K_1 R_1) \notag\\
  &= K_{2 \mathrm{a}} R_1^2 - K_1 R_1 + \cfrac{K_1^2}{4 K_{2 \mathrm{a}}}.
\end{align}
%Thirdly, we obtain the second derivatives of the potential at three extremal points:
The second derivatives of the potential at three extremal points are
\begin{align}
  \partial_{\theta}^2 U(\theta_{\max}, R_1) &= \cfrac{K_1^2}{2 K_{2 \mathrm{a}}} - 2 K_{2 \mathrm{a}} R_1^2,\\
  \partial_{\theta}^2 U(\theta_{\min 1}, R_1) &= \partial_{\theta}^2 U(0) = 2 K_{2 \mathrm{a}} R_1^2 + K_1 R_1,\\
  \partial_{\theta}^2 U(\theta_{\min 2}, R_1) &= \partial_{\theta}^2 U(\pi, R_1) = 2 K_{2 \mathrm{a}} R_1^2 - K_1 R_1.
\end{align}

We assume that the noise is sufficiently weak compared to $\Delta U$, i.e., $D \ll \Delta U$.
We also assume that $R_1$ evolves sufficiently slowly. Then, we can expect that the phase distribution is well approximated to 
\begin{align}
  P(\theta,t)=\eta(t) \delta(\theta) + (1 - \eta(t)) \delta(\theta - \pi).
  \label{delta function}
\end{align}
In this approximation, each oscillator takes the phase either $0$ or $\pi$.
By defining $H$ and $H^*$ as the states in which the phase of an oscillator has the phases $0$ and $\pi$, respectively, the transition process is schematically described as
\begin{align}
  \label{seq: reaction}
  H \overset{k_+}{\underset{k_-}{\rightleftharpoons}} H^*,
\end{align}
where $k_\pm$ are the transition rates given by
% We obtainz
\begin{align}
  k_+ (R_1) %&= k_+ = k_- \notag\\
  &= 2 \cfrac{\sqrt{|\partial_{\theta}^2 U(\theta_{\min 1}, R_1)\partial_{\theta}^2 U(\theta_{\max}, R_1)}|}{2\pi} \exp\left(- \cfrac{\Delta U_1}{D}\right), \label{k+}\\
  k_- (R_1) %&= k_+ = k_- \notag\\
  &= 2 \cfrac{\sqrt{|\partial_{\theta}^2 U(\theta_{\min 2}, R_1)\partial_{\theta}^2 U(\theta_{\max}, R_1)}|}{2\pi} \exp\left(- \cfrac{\Delta U_2}{D}\right). \label{k-}
\end{align}
Substituting the obtained expressions into Eqs.~\eqref{k+} and \eqref{k-}, we obtain
\begin{align}
  \label{seq: rate}
  k_+ (R_1) &= \cfrac{1}{\pi} \sqrt{(2 K_{2 \mathrm{a}} R_1^2 + K_1 R_1)\left|2 K_{2 \mathrm{a}} R_1^2 - \cfrac{K_1^2}{2 K_{2 \mathrm{a}}} \right|} \exp \left(- \cfrac{1}{D} \left(K_{2 \mathrm{a}} R_1^2 + K_1 R_1 + \cfrac{K_1^2}{4 K_{2 \mathrm{a}}} \right) \right),\\
  k_- (R_1) &= \cfrac{1}{\pi} \sqrt{(2 K_{2 \mathrm{a}} R_1^2 - K_1 R_1)\left|2 K_{2 \mathrm{a}} R_1^2 - \cfrac{K_1^2}{2 K_{2 \mathrm{a}}} \right|} \exp \left(- \cfrac{1}{D} \left(K_{2 \mathrm{a}} R_1^2 - K_1 R_1 + \cfrac{K_1^2}{4 K_{2 \mathrm{a}}} \right) \right).
\end{align}

\begin{figure}[htb]
 \begin{tabular}{cc}
   \begin{minipage}[t]{5.0cm}
     \centering
     \includegraphics[height = 4.0cm]{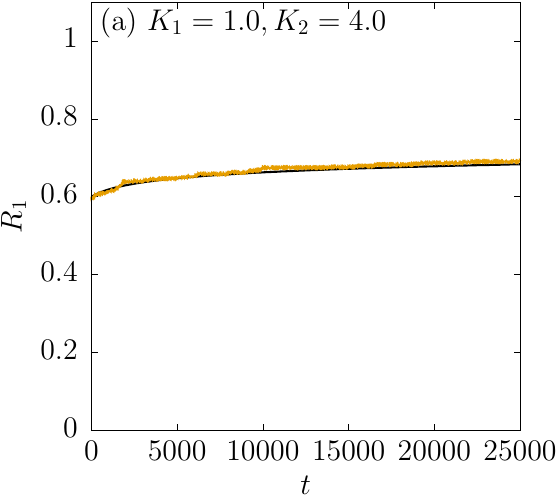}
   \end{minipage} &
   \begin{minipage}[t]{5.0cm}
     \centering
     \includegraphics[height = 4.0cm]{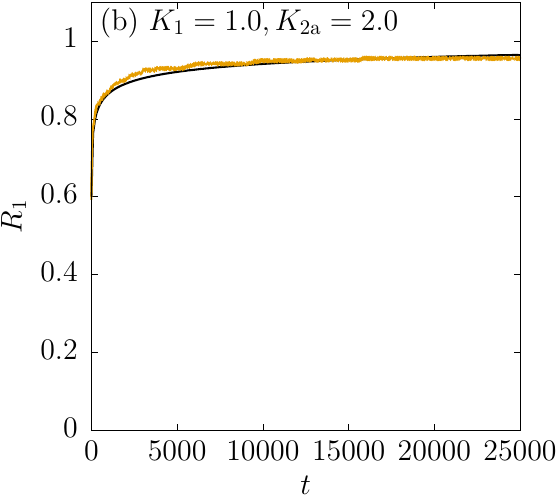}
   \end{minipage}
 \end{tabular}
 \caption{Comparison between the time evolutions of $R_1$ and the simulations of $N = 10^3$ oscillators. 
 The black lines represent the numerical simulations of $R_1$ following Eq. \eqref{seq: R_1 Kramers 2}, 
 while the orange lines show the dynamics of $R_1$, which are averaged over $10^3$ oscillators governed by Eq. \eqref{eq: model 1}.}
 \label{sfig: R_1 Kramers}
\end{figure}

We find that the time evolution of $R_1$ is
\begin{align}
  \dot{R_1} &= - k_+ (1 + R_1) + k_- (1 - R_1)\\
%   &= - \cfrac{1}{\pi} \sqrt{(2 K_{2 \mathrm{a}} R_1^2 + K_1 R_1)\left|2 K_{2 \mathrm{a}} R_1^2 - \cfrac{K_1^2}{2 K_{2 \mathrm{a}}} \right|}
%   \exp \left(- \cfrac{1}{D} \left(K_{2 \mathrm{a}} R_1^2 + K_1 R_1 + \cfrac{K_1^2}{4 K_{2 \mathrm{a}}} \right) \right) (1 + R_1) \notag\\
%   &+ \cfrac{1}{\pi} \sqrt{(2 K_{2 \mathrm{a}} R_1^2 - K_1 R_1)\left|2 K_{2 \mathrm{a}} R_1^2 - \cfrac{K_1^2}{2 K_{2 \mathrm{a}}} \right|}
% \exp \left(- \cfrac{1}{D} \left(K_{2 \mathrm{a}} R_1^2 - K_1 R_1 + \cfrac{K_1^2}{4 K_{2 \mathrm{a}}} \right) \right) (1 - R_1) \notag\\
  &= - \cfrac{1}{\pi} \sqrt{(2 K_{2 \mathrm{a}} R_1^2 + K_1 R_1)\left(2 K_{2 \mathrm{a}} R_1^2 - \cfrac{K_1^2}{2 K_{2 \mathrm{a}}} \right)}
  \exp \left(- \cfrac{1}{D} \left(K_{2 \mathrm{a}} R_1^2 + K_1 R_1 + \cfrac{K_1^2}{4 K_{2 \mathrm{a}}} \right) \right) (1 + R_1) \notag\\
  &+ \cfrac{1}{\pi} \sqrt{(2 K_{2 \mathrm{a}} R_1^2 - K_1 R_1)\left(2 K_{2 \mathrm{a}} R_1^2 - \cfrac{K_1^2}{2 K_{2 \mathrm{a}}} \right)}
\exp \left(- \cfrac{1}{D} \left(K_{2 \mathrm{a}} R_1^2 - K_1 R_1 + \cfrac{K_1^2}{4 K_{2 \mathrm{a}}} \right) \right) (1 - R_1).
  \label{seq: R_1 Kramers 2}
\end{align}
% Figure \ref{sfig: R_1 Kramers} compares the dynamics of $R_1$ given by \eqref{seq: R_1 Kramers 2} to those of the simulations of $N = 10^3$ oscillators governed by Eq. \eqref{eq: model 1}, which are in an excellent agreement. 
Figure \ref{sfig: R_1 Kramers} compares the dynamics of $R_1$ given by \eqref{seq: R_1 Kramers 2} to those of the simulations of $N = 10^3$ oscillators governed by Eq. \eqref{eq: model 1}.
This comparison is made in a regime where the synchronized states are stable, showing excellent agreement.

\begin{figure}[htb]
    \centering
 \includegraphics[height = 5cm]{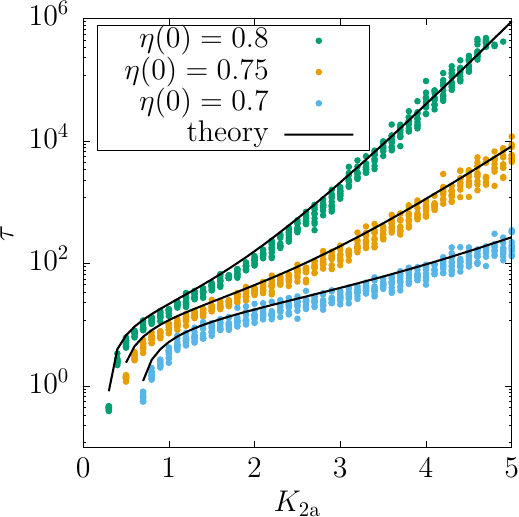}
 \caption{Lifetime of the synchronized states versus $K_{2 \mathrm{a}}$. Simulation results of Eq.~\eqref{eq: model 1} with $N = 10^3$ oscillators and the theoretical prediction, given by Eq.~\eqref{eq: T thre}, are shown by colored plots and solid lines, respectively. 
 We fix $K_1=0$, $K_{2 \mathrm{b}} = 0$ and $D=0.1$.}
    \label{fig: tau}
\end{figure}

To specifically focus on the lifetime of the synchronized state, we now consider the simplified case of $K_1 = 0$, where the Eq. \eqref{seq: R_1 Kramers 2} reduces to the following form:

\begin{align}
  \dot{R_1}(t) = - \cfrac{4 K_{2 \mathrm{a}} R_1^3}{\pi} \exp \left(- \cfrac{K_{2 \mathrm{a}} R_1^2}{D}\right).
  \label{eq: time evolution of R_1}
\end{align}
Because $\dot R_1<0$ for $R_1>0$ and $\dot R_1=0$ for $R_1=0$, $R_1=0$ is the global attractor.
However, as the term $\exp (- K_{2 \mathrm{a}} R_1^2/D)$ is vanishingly small for
%$R_1 \gg \sqrt{\frac{D}{K_{2 \mathrm{a}}}}\equiv R_\mathrm{thre}$,
$R_1 \gg \sqrt{\frac{D}{K_{2 \mathrm{a}}}}\equiv R^*$,
the relaxation to $R_1=0$ is extremely slow if $R_1(0)\equiv \hat R \gg R^*$.
We define the lifetime $\tau$ of the synchronized state as the time for which $R_1$ varies from
$\hat R$ to $R^*$.
Because $\tau=\int_{0}^{\tau}dt = \int_{\hat R}^{R^*}\frac{dt}{dR_1}dR_1$, we obtain
\begin{align}
  \tau %
  &= \int_{R^*}^{\hat R} \cfrac{\pi \exp \left(\frac{K_{2 \mathrm{a}} R_1^2}{D}\right)}{4 K_{2 \mathrm{a}} R_1^3}\  \mathrm{d}R_1.
  \label{eq: T thre}
\end{align}
This integral can only be computed numerically. 
Because the evolution of $R_1(t)$ is very slow until $R_1$ reaches $R^*$, a rough estimate of $\tau$ 
can be given by setting $R_1(t)=\hat R$ in Eq.~\eqref{eq: T thre}, giving rise to
\begin{align}
 \tau \sim {\rm exp} \left(\frac{K_{2 \mathrm{a}} {\hat R}^2}{D}\right),
 \label{tau_rough}
\end{align}
where the coefficient, including the factor
%$\frac{R_0-R_\mathrm{thre}}{R_\mathrm{thre}^3}$
$\frac{\hat R-R^*}{\hat R^{3}}$
is omitted. This estimation indicates that $\tau$ approximately exponentially increases with $K_{2 \mathrm{a}}$.
To verify our theory, in Fig.~\ref{fig: tau},
we compare our theoretical estimations, given by Eqs.~\eqref{eq: T thre} and \eqref{tau_rough}, to the lifetime obtained from the direct simulations of Eq.~\eqref{eq: model 1}.
%The simulation setup is identical to that illustrated in Fig.~\ref{fig: dynamics of R_1}, and
The lifetime is given as the time at which $R_1$ passes $R^*$ for the first time. 
Note that $\hat{R} = |2 \eta(0) - 1|$, where $\eta(0)$ is the parameter for the initial distribution. We observe that Eq.~\eqref{eq: T thre} is in reasonable agreement with the simulation data. 
% We also observe that the lifetime indeed increases approximately exponentially with $K_{2 \mathrm{a}}$, as predicted by Eq.~\eqref{tau_rough}.
We also observe that the lifetime indeed increases approximately exponentially with $K_{2 \mathrm{a}}$, as predicted by Eq.~\eqref{tau_rough}, supporting the robustness of this phenomenon (see Appendix \ref{sec: Appendix D}).
% Moreover, qualitatively similar results are obtained for nonvanishing $K_1$ or $K_{2 \mathrm{a}}$ values. Please refer to the Appendix \ref{sec: Appendix D}.
%As detailed in \cite{Note1}, the closed equation for $R_1$ may also be obtained for $K_1>0$.

%and it approximately reproduces $R_1(t)$ trajectories obtained in Eq.~\eqref{eq: model 1}.

\section{Conclusion}
% {\em Conclusion.}
In this study, we explored 
a large population of noisy oscillators with one- and two-simplicial interactions.
%Specifically, we demonstrated that dynamical noise, no matter how weak, will extinguish steady synchronized states when only the type-a two-simplicial interaction exists.
We demonstrated that dynamical noise, regardless of its strength, erodes the synchronized states when the one-simplex and type-b two-simplex interactions are absent or sufficiently weak. 
However, the lifetime of the synchronized state is prolonged, increasing exponentially with the strength of the type-a two-simplex interaction.
Note that the characteristic time scales of individual units, given by the inverse of interaction and the noise strength, are small in our setting. For example, the time scale of the diffusion process is
$\frac{1}{\sqrt{D}}$, which is approximately 3. Compared to them, the erosion process is extremely slow. 
Thus, the system can be considered synchronous or asynchronous depending on the time scale of the observation. We also emphasize that the erosion is robust, which is observed in the presence of different types of interactions and frequency heterogeneity. 
%We name this slow decay process as the noise erosion of the synchronized state.
%Nevertheless, the decay of the synchronized state is slow. Specifically, it has a characteristic time scale that increases 
%In addition, this erosion of synchronization is robust with one-simplex interaction or type-b two-simplex interaction if they are weak to a certain extent.
%However, synchronized states composed of two clusters persist for extended periods.
%Our study uncovered the processes of synchronization and desynchronization of noisy oscillators in higher-order networks and is expected to provide insight into the design and control principles of such systems.
The proposed theory, which can handle mixed systems of one-simplex and two types of two-simplex interactions, clarified the bifurcation structure and transient dynamics. 
%Our study provides insights into the design and control principles of oscillator assemblies in higher-order networks.

%This feature could translate to future applications. In the context of memory storage, a cluster can be considered as a state in which the system holds some information. Since the lifetime of a cluster state is determined by the initial ratio of clusters, it may be possible to design a self-organizing system that can measure time and also pre-set the time to retain information.

%Finally, we discuss the limitations and future directions of this study.
%We have considered only the all-to-all coupling case. 
%Although such a network is mathematically tractable, its applicability to real-world systems would be limited. 
%Extension to complex networks is vital. 
% Furthermore, to comprehensively understand the nature of multi-simplex interaction, 
% it is essential to explore another generic coupling type, given as $\sin(\theta_j-\theta_k-\theta_m)$ instead of $\sin(2\theta_j+\theta_k-2\theta_m)$ in Eq.~\eqref{eq: model 1} \cite{ashwin2016hopf,Skardal2020-wq}. 
% The noise effect on such a system is an interesting open problem.

\bibliography{citations}

\appendix

%\section{Relation between $\Theta_1$ and $\Theta_2$}
\section{Time evolution of $\Theta_1$ and $\Theta_2$}
\label{sec: Appendix A}
\begin{figure}[htb]
  \begin{tabular}{cc}
    \begin{minipage}[t]{5.0cm}
      \centering
      \includegraphics[height = 4.0cm]{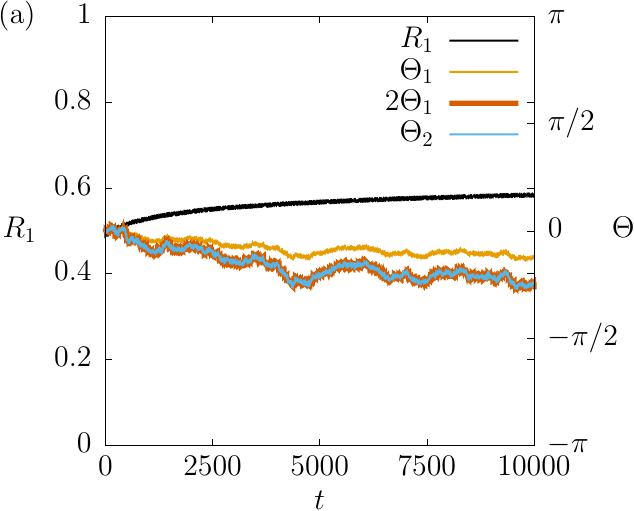}
    \end{minipage} &
    \hspace{0.5cm}
    \begin{minipage}[t]{5.0cm}
      \centering
      \includegraphics[height = 4.0cm]{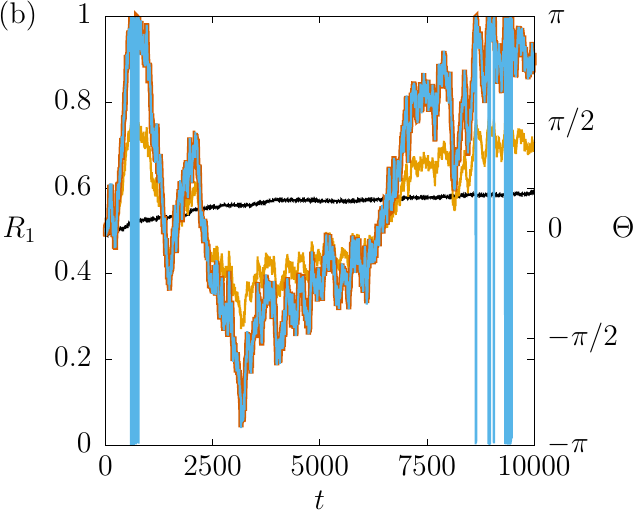}
    \end{minipage}
  \end{tabular}
  \caption{Time evolution of
%the order parameters
order parameter $R_1$ and mean phases $\Theta_1$ and $\Theta_2$ for $K_{2 \mathrm{b}}$ = 0.3, $K_{2 \mathrm{a}} = 4.0$, and $D = 0.1$. (a) $N = 10^5$ and (b) $N = 10^3$.
%oscillators.
}
  \label{sfig: Theta}
 \end{figure}

%In this section, we prove that $\Theta_2 = 2 \Theta_1$ is a steady solution to Eq. \eqref{eq: FPE}.
Figure \ref{sfig: Theta}, which illustrates the results of direct simulations of Eq. \eqref{eq: model 1}, shows that 
the trajectories of $2\Theta_1$ are the same as those of $\Theta_2$ in synchronized states.
Moreover, $\Theta_1$ and $\Theta_2$ evolve more slowly as the number of oscillators increases. Therefore, we expect that 
$\dot{\Theta}_l \to 0$ as $N \to \infty$.

\section{Numerical analysis of self-consistency equation}
\label{sec: Appendix B}
\begin{figure}[htb]
  \centering
  \includegraphics[height = 4.5cm]{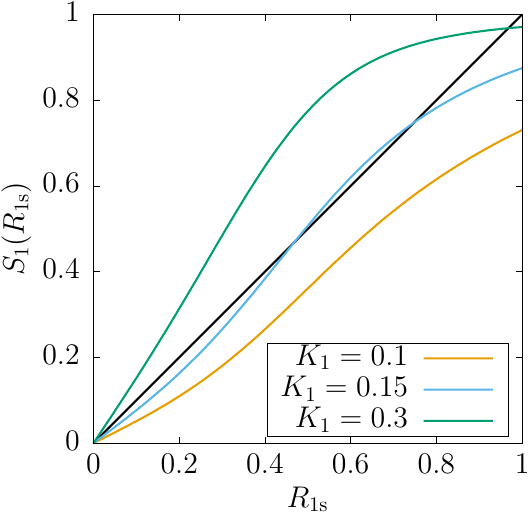}
  \caption{Typical behavior of the self-consistency equation \eqref{eq: selfconsistency 1} for $K_{2 \mathrm{a}} = 1.0$ and $K_{2 \mathrm{b}} = 0$ with varying $K_1$.}
  \label{sfig: selfconsistency}
\end{figure}

In this section, we briefly
%present our procedure of self-consistency analysis. 
illustrate a numerical analysis of the self-consistency equation, given by Eq. \eqref{eq: selfconsistency 1}.
For $K_{\rm 2b} = 0$, we need to solve Eq. \eqref{eq: selfconsistency 1} only for $l=1$ because $R_2$ is not involved. Figure \ref{sfig: selfconsistency} shows $S_1 (R_{1 \mathrm{s}})$ vs $R_{1 \mathrm{s}}$ and $R_{1 \mathrm{s}}$ vs $R_{1 \mathrm{s}}$, and its concurrent points can easily be identified using, e.g., a bisection method.

\begin{figure}[htb]
  \centering
  \begin{tabular}{cc}
    \begin{minipage}[t]{5.0cm}
      \centering
      \includegraphics[height = 4.0cm]{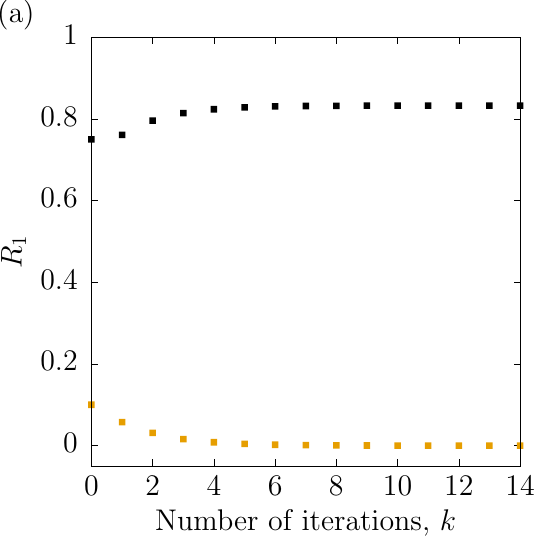}
    \end{minipage}
    \begin{minipage}[t]{5.0cm}
      \centering
      \includegraphics[height = 4.0cm]{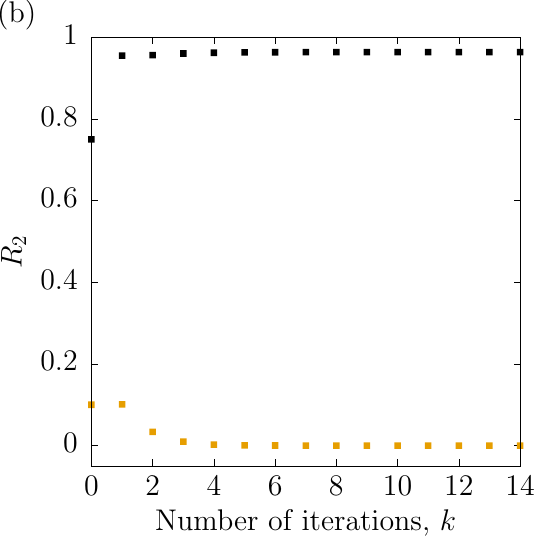}
    \end{minipage}
  \end{tabular}
  \caption{Visualization of the iteration (\ref{seq: selfconsistency iteration 1}, \ref{seq: selfconsistency iteration 2}) for $K_1 = 0.1$, $K_{2 \mathrm{a}} = 4.0$, $K_{2 \mathrm{b}} = 0.05$ and $D = 0.1$. The black and orange symbols present the time courses of $R_1^k$ and $R_2^k$ for $(R_1^0,R_2^0)=(0.75,0.75)$ and $(R_1^0,R_2^0)=(0.1,0.1)$, respectively.}
  \label{sfig: selfconsistency iteration}
\end{figure}

For $K_{\rm 2b} \neq 0$, we use the following iterative method:
%obtain an upper branch in the synchronization diagrams such as Figs. \ref{fig:phase diagram}(c) and (d):
\begin{align}
  \label{seq: selfconsistency iteration 1}
  R_{1}^{k+1} &= S_1 (R_1^k, R_2^k),\\
  \label{seq: selfconsistency iteration 2}
  R_{2}^{k+1} &= S_2 (R_1^k, R_2^k).
\end{align}
%When $K_{2 \mathrm{b}} = 0$, the iteration is further simplified into the equation for $R_1$. 
%Figure \ref{sfig: selfconsistency} illustrates typical behavior of $S_1 (R_{1 \mathrm{s}})$, indicating that the steady state of $R_{1 \mathrm{s}}$ bifurcates as $K_1$ and $K_{2 \mathrm{a}}$ vary.
We expect that $(R_{1 \mathrm{s}},R_{2 \mathrm{s}})$ corresponding to a stable steady state of Eq.~\eqref{seq: FPE reduced} is obtained for an appropriate initial condition.
In Fig. \ref{sfig: selfconsistency iteration}, we plot the values of $R_l^k$ at each step of iteration in the bistable region.
We observe that either the desynchronized or synchronized state is obtained depending on the initial condition.
%a convergence value depends on an initial value.

\section{Phase diagram in the $(K_1, K_{2 \mathrm{a}})$ plane for nonvanishing $K_{2 \mathrm{b}}$}
\label{sec: Appendix C}
\begin{figure}[htb]
  \centering
  \includegraphics[height = 4.5cm]{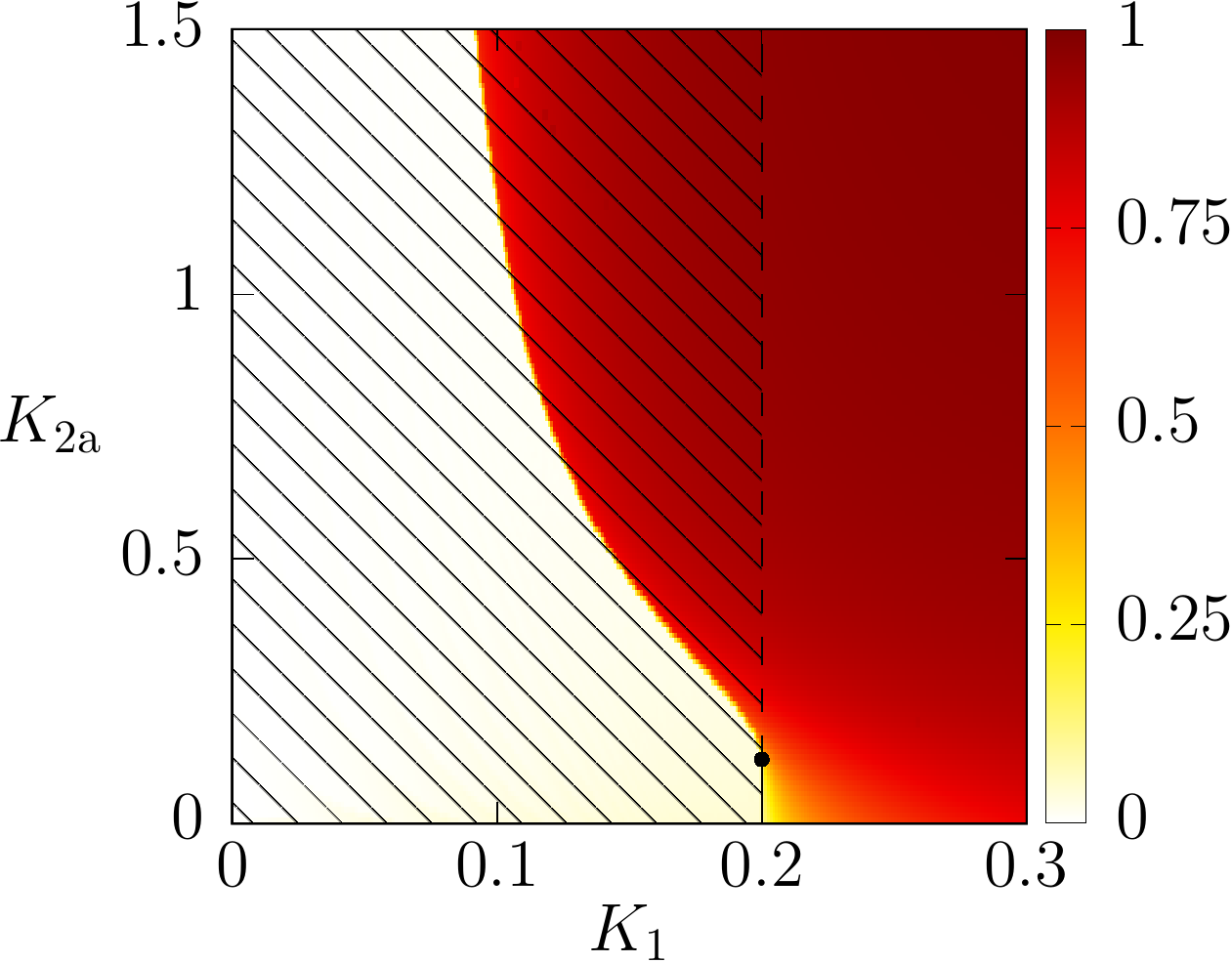}
  \caption{Phase diagram in the $(K_1, K_{2 \mathrm{a}})$ plane for $K_{2 \mathrm{b}} = 0.05$ and $D = 0.1$.
  The color scale describes the $R_1$ value of the stable synchronized state. The hatched region denotes the region in which the desynchronized state is stable.}
  \label{sfig: phase diagram}
\end{figure}

Figure \ref{sfig: phase diagram} is a phase diagram in $(K_1, K_{2 \mathrm{a}})$ plane for $K_{2 \mathrm{b}} = 0.05$. 
The point in Fig. \ref{sfig: phase diagram} that marks the bifurcation type switch at $K_1 = 2D$, as described by Eq.~\eqref{eq: bifurcation}.

\section{Lifetime of synchronization for $K_1 > 0$ or $K_{2 \mathrm{b}} > 0$}
\label{sec: Appendix D}
\begin{figure}[htb]
  \centering
  \begin{tabular}{cc}
    \begin{minipage}[t]{5.0cm}
      \centering
      \includegraphics[height = 4.0cm]{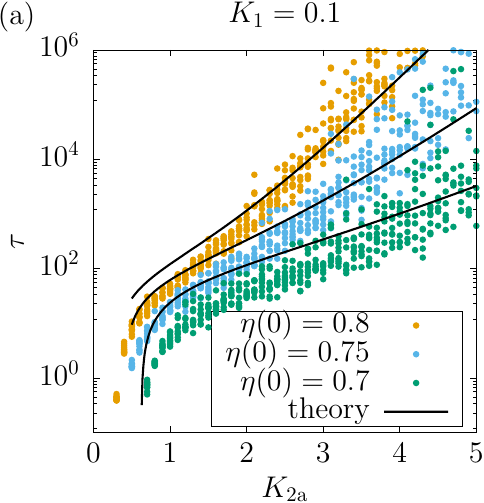}
    \end{minipage} &
    \begin{minipage}[t]{5.0cm}
      \centering
      \includegraphics[height = 4.0cm]{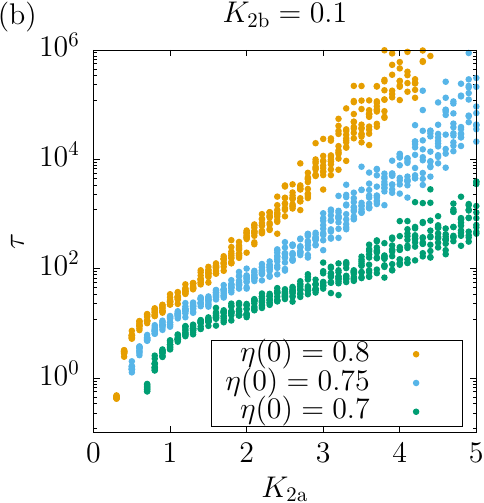}
    \end{minipage}
  \end{tabular}
  \caption{Lifetime of synchronization versus $K_{2 \mathrm{a}}$ for (a) $K_1 = 0.1$ and $K_{2 \mathrm{b}} = 0$, and (b) $K_1 = 0$ and $K_{2 \mathrm{b}} = 0.1$. Results of simulations for $N = 10^3$ and $D = 0.1$.}
  \label{sfig: tau}
\end{figure}

In Fig. \ref{sfig: tau}, we plot the lifetime of synchronization when the one-simplex or type-b interaction is present ($K_1 = 0.1$ or $K_{2 \mathrm{b}} = 0.1$).
In this case, no synchronized states are stable for any $K_{2 \mathrm{a}}$.
We find that the lifetime has an exponential dependence on $K_{2 \mathrm{a}}$ as well as when $K_1 = 0$ and $K_{2 \mathrm{b}} = 0$. 

\end{document}